\documentstyle[lprocl,11pt,epsf,wrapfig]{article}

\def\be{\begin{equation}}
\def\ee{\end{equation}}
\def\bea{\begin{eqnarray}}
\def\eea{\end{eqnarray}}
\def\gr{$\gamma$-ray }
\def\grs{$\gamma$-rays }
\def\ga{gamma ray astronomy }

\begin{document}

\title{VERY HIGH ENERGY PHOTONS IN UNIVERSE}

\author{F. A. AHARONIAN}

\address{Max-Planck-Institut f\"ur Kernphysik, Postfach 10 39 80\\
D-69029 Heidelberg, Germany\\E-mail: aharon@fel.mpi-hd.mpg.de}

\maketitle\abstracts{I discuss the recent exciting results
of discovery of TeV photons from several galactic and extragalactic 
sources,  and  highlight some objectives of VHE
gamma ray astronomy  that can be achieved in the near future
by means of new generation ground-based detectors.}

\section{Introduction}
The branch of high energy astrophysics that studies the sky in 
energetic \gr photons -- {\it \ga} -- is destined to play a crucial 
role in exploration of nonthermal phenomena in the Universe in 
their most violent forms. The great potential of the discipline
is provided by two key circumstances:
(1) copious production of \grs in a variety of  
galactic and  extragalactic sources, and
(2) availability of effective space- and ground-based 
techniques of detection of cosmic 
$\gamma$-radiation in very broad energy range from several 
hundreds keV to $\geq 10 \, \rm TeV$.
This allows rather impressive coverage of a 
variety of `hot' topics of the modern astrophysics 
and cosmology, such as 

\begin{itemize}

\item Origin of galactic and extragalactic cosmic rays

\item Acceleration and radiation processes at extreme astrophysical 
conditions, in particular  in the pulsar magnetospheres,  
relativistic jets of active galactic nuclei, {\it etc.}

\item Physics and astrophysics of relativistic objects -- 
neutron stars and black holes

\item Nature of enigmatic astrophysical phenomena like the 
$\gamma$-ray bursts

\item Cosmological issues connected with the level of 
diffuse  photon fluxes and  magnetic fields
on large extragalactic scales  and their evolution in time; 
independent measurements of the Hubble constant; search for
dark matter in the form of WIMPs through their characteristic
annihilation radiation; test of the non-acceleration (`top-down')
scenarios of production of particles observed in cosmic rays
above $10^{20} \, \rm eV$, {\it etc.} 

\end{itemize}

The recent exciting results~\cite{egret} from the Energetic Gamma Ray Experiment 
Telescope (EGRET) aboard the Compton Gamma Ray Observatory (GRO)   
start to confirm these prime motivations of 
gamma ray astronomy. Indeed, many typical 
representatives of different
galactic and extragalactic source populations, in particular pulsars
(Crab, Vela), supernovae remnants ($\gamma$~Cygni, IC 443),
plerions (Crab Nebula), giant molecular clouds/star formation regions
(Orion, $\rho$~Ophiuchus), quasars (3C~273, 3C~279), {\it etc.},  
which were predicted
as potential GeV gamma ray emitters, now are among of more than 300 
\gr sources detected by EGRET. 
This success  has elevated the field to the level of  
truly  {\it observational} discipline.  At the same time, the nature of 
most of the EGRET sources remains unknown. Moreover, the origin of 
$\gamma$-radiation from even firmly identified objects is poorly understood.
This clearly justifies the future gamma ray missions
with new generation detectors like the {\bf G}amma-ray 
{\bf L}arge {\bf A}rea {\bf S}pace {\bf T}elescope (GLAST)~\cite{glast}.

The GLAST, with its large (almost $2 \pi$ steradian) field of view 
and superior flux sensitivity,  
$J_{\rm min} \sim 10^{-10} \, \rm ph/cm^2 s$ above 1 GeV, seems to be
an ideally designed instrument for  deep surveys of the sky in 
GeV $\gamma$-rays with an ambitious aim to do 
{\it ``$\gamma$-ray astronomy 
with thousands of sources}''~\cite{dingus}.    
The significant advances in the detection area, angular resolution and field 
of view of GLAST as compared to EGRET  provide very strong, 
by a factor of 100, reduction  
in the sensitivity threshold  (see Fig.~1) that ultimately could result in
dramatic increase of number of $\gamma$-ray sources. In particular, 
it is expected  that more than 1000 (!)  active galactic 
nuclei will be  detected  in GeV $\gamma$-rays~\cite{dingus}. 
A {\it detailed  study} of the spatial, temporal, and spectral 
characteristics  of already  identified  \gr sources 
is the second major objective of the GLAST. 
Also, since most of the EGRET 
sources do not exhibit spectral cutoffs in the 1-10 GeV region, 
the extension of the study into the unexplored region beyond 
10 GeV is  believed to be another important issue for the GLAST.   
Meanwhile, the area limitations of space-borne detectors 
compels the region of very high 
energy (VHE) photons above 100 GeV to remain (except for the 
specific topics related to the  diffuse galactic and  extragalactic  
backgrounds) 
the  {\it domain of ground-based gamma ray astronomy}.  

\section{The Status of VHE Gamma Ray Astronomy}

The basic idea of study of the cosmic $\gamma$-radiation by means of
ground-based instruments  is linked to the detection of secondary 
particles and/or the Cherenkov radiation  of 
electromagnetic cascades produced in the Earth's atmosphere 
by very energetic \gr photons~\cite{cronin_annrev93}.
The history of the ground-based  \ga dates 
back to early sixties, but unfortunately the development of the 
field over  many years  was rather slow and  controversial.  
One of the likely reasons for the slow progress could be 
the fact that the  `TeV community' has only recently 
realized that the proper {\it detection/identification} of primary \grs 
requires more sensitive and sophisticated methods and detectors than 
the ones used in the past in traditional cosmic ray  experiments. 
Indeed, the exploitation of the so called {\it imaging atmospheric 
Cherenkov telescope} (IACT) technique 
was soon rewarded  by the $9 \sigma$ detection of TeV \grs 
from the Crab Nebula~\cite{whipple_crab89}. 
This seminal result of the Whipple collaboration    
published in 1989  perhaps could be registered as the  first 
reliable observation of \grs from a celestial object
obtained with a ground-based detector.

%fig.1 (crab/sensitivities)
\begin{figure}[t]
%\epsfxsize=12  cm
%\epsfysize=11.9 cm
%\epsffile[8 415 560 760]{fg1.ps}
\vspace{8.  cm}
\includegraphics{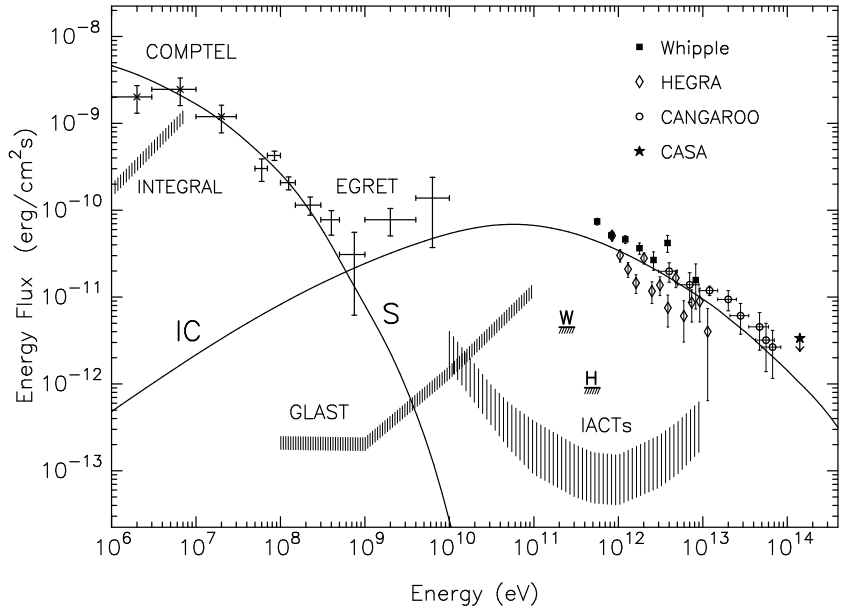}
\caption{Sensitivities of satellite-borne
and ground-based gamma ray detectors at the $5 \sigma$ detection threshold
compared with the flux of the Crab Nebula
measured at MeV/GeV  (COMPTEL/EGRET) and at TeV energies (only three recent TeV
measurements by the Whipple (1995/1996) [4],
HEGRA (1996/1997) [5], and CANGAROO
(1994/96) [6], as well as the flux limit
set by CASA/MIA [7] are shown).
Two solid curves correspond to the calculated [8]
synchrotron (S) and inverse Compton (IC) contributions to the Crab
spectrum.  The detector sensitivities correspond to the following
conditions: INTEGRAL - exposure time $10^6 \, \rm s$ [9],
GLAST -- 1 year all-sky survey [2],
IACTs arrays -- 100 h `on-source' observations. 
The current sensitivities of the Whipple 10 m telescope and the
HEGRA IACT system for 100~h observation time are also shown at the relevant energy thresholds.
The sensitivities of the COMPTEL and EGRET approximately 
correspond  to the fluxes of the Crab Nebula 
detected by these instruments.}
\end{figure}

The high {\it detection rate}, 
the ability of effective {\it separation} of electromagnetic and 
hadronic showers, and the good accuracy of {\it reconstruction}
of air shower parameters are three remarkable features of the
IACT technique~\cite{Caw/Weekes_96,AhAk_annrev97}.
The recent reports about detection of 
VHE \grs from several objects by nine  groups
running imaging telescopes both in the northern (Whipple,
HEGRA, CAT, Telescope Array, CrAO, SHALON, TACTIC) and southern
(CANGAROO, Durham) hemispheres confirm the early expectations
concerning the potential of the 
technique~\cite{Weekes_Turver,Stepanian,Hillas_LaJolla}, 
and provide a solid basis for the ground-based \ga~\cite{GRO_Rev}.

\subsection{TeV Energy Domain}

Presently, four celestial objects are  unambiguously identified  
as TeV \gr sources: Crab, PSR 1706-44,
Mrk 421, and Mrk 501. These objects  have been detected by two or more 
independent groups with $\geq 10$ standard deviation statistical
significance, and represent, most probably, two  well recognized 
nonthermal source populations -- pulsar driven nebulae (or plerions) 
and active galactic nuclei (AGN)~\cite{AhAk_annrev97,GRO_Rev}.
Actually, three more candidates, detected with statistical significance
of about $6 \sigma$ or more,  should be 
added to this viable list of TeV \gr sources:~ a pulsar driven nebula -- 
Vela~\cite{cang_vela}, a
shell-type supernova remnant (SNR) --  SN~1006~\cite{cang_sn1006}, 
and a nearby AGN -- 1ES 2344+514~\cite{whipple_1es}.

And finally, one has to mention some intriguing indications of episodic 
TeV signals from X-ray binaries. Two famous representatives of this population
of compact galactic sources, Cyg X-3 and Her X-1 dominated VHE and UHE
\ga in the 80's, and played a crucial role in the renewed interest 
in ground-based \gr observations~\cite{Weekes_Pysrep88,Fegan_ad90}.
But unfortunately almost all the early claims about detection of 
\gr signals from these objects both at TeV and PeV energies, were not 
confirmed by later, more sensitive observations~\cite{cronin_annrev93}. 
Although one cannot exclude a long-term variability of these sources, 
many experts treat the old results with a certain skepticism. Nevertheless, 
perhaps it is premature to draw final conclusions, especially if one takes
into account the recent claims about detection of TeV signals from  
the cataclysmic variable AE Aquarii~\cite{potch_AEaq,durham_AEaq} 
galactic black hole candidate  GRS1915+105  
(`microquasar')~\cite{Perpig_hegra}, 
X-ray binaries Vela X-1~\cite{potch_VelaX-1} and Cen X-3~\cite{durham_cenx3}. 
Interestingly, a \gr signal from Cen X-3 recently 
has been claimed to be detected also at GeV energies by 
EGRET~\cite{egret_cenx3}.
Although the statistical significance and/or 
systematic uncertainties of these results 
do not match the standards set by  current IACTs, they indicate a necessity 
of monitoring these sources
by new generation of imaging Cherenkov telescopes.   

Along with above mentioned  `firmly-detected' and  `possible-candidate'
TeV sources, there are several tens of objects that have been observed 
without finding significant evidence for a TeV signal. 
In some cases the flux upper limits,  e.g. from the Crab and Geminga pulsars,
and especially from shell type SNRs $\gamma$ Cygni and IC 443, 
have begun to have  important  theoretical 
consequences~\cite{buckley_SNR,SNRrev_GRO4,vlk_SNR}. Also, several interesting
searches for TeV \gr transients of different origins (e.g due to
evaporation of primordial black holes (PBH), or possible continuation of 
the spectra of low-energy gamma ray bursts (GRBs)) 
have been made by Cherenkov telescopes and air shower arrays.
Meaningful upper limits have already been set to the PBH density, and
upper limits to the delayed (or extended) TeV emission 
were derived from the immediate monitoring of GRBs 
using the BACODINE information network (see e.g. [17] and references
therein).

The implications of above mentioned results  in the general  context of 
scientific  objectives of VHE  \ga are discussed below (Sec.~4).  

\subsection{PeV Energy Domain}

Despite intensive efforts in the early 90's that were motivated by 
early claims about detection of PeV \grs from Cyg X-3, all sky surveys
by sensitive air shower detectors CASA-MIA, Cygnus, EAS-TOP, 
HEGRA, SPASE, Tibet {\it etc.}, could not detect point sources of ultra high energy 
(UHE) \grs above 20 TeV~\cite{Cronin_Rome}.
Although very important, and in some cases astrophysically meaningful,
e.g. the  upper limits on the diffuse galactic and extragalactic \gr fluxes
set by the CASA/MIA~\cite{casa_GALdiffuse,casa_EXGdiffuse}, these results hardly 
can change an impression that the early promise of UHE  \ga has not been 
fulfilled.
%fig.2
\begin{figure}[t]
%\epsfxsize=12 cm
%\epsfysize=11.9 cm
%\epsffile[40 150 565 695]{fg2.ps}
\vspace{8.3  cm}
\includegraphics{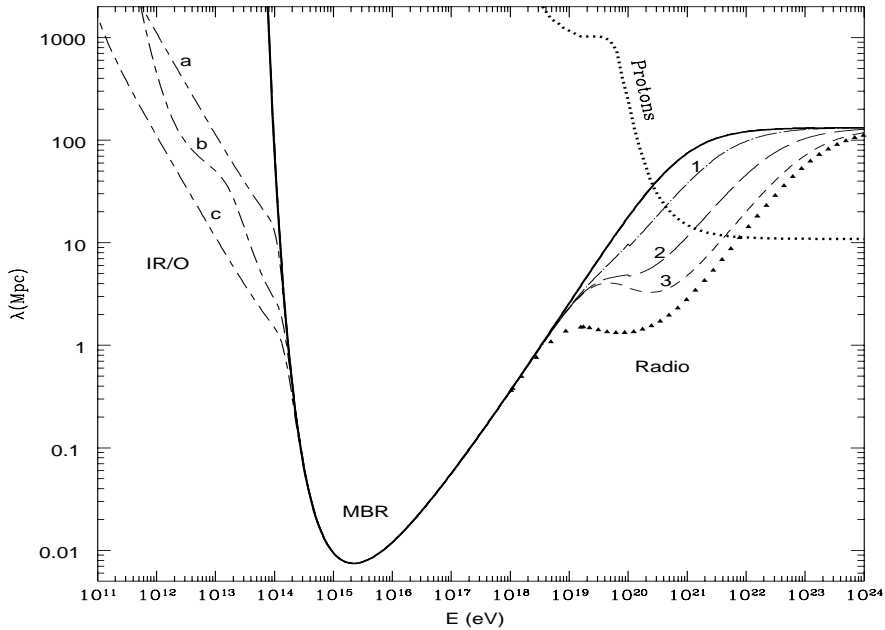}
\caption{
The attenuation lengths of protons (dotted curve) and
$\gamma$-rays in the intergalactic space calculated [37] for the 
{\it current} density of DEBRA.  
The \gr pathlength  at $E \leq 10^{14} \, \rm eV$ and
$E \geq 10^{19} \, \rm eV$ presented by curves a;b;c and 1;2;3, respectively,
reflect the uncertainty of the level of the 
photon background at the IR/O ($\lambda \sim 0.1-100 \, \mu m$) and
radio ($\nu \sim 1-10 \, \rm MHz$) wavelengths. Meanwhile the 
$\gamma$-ray pathlength in the energy region from $10^{14} \, \rm eV$
to $10^{19} \, \rm eV$ is dominated by the interactions with 
2.7 K MBR.}
\end{figure}
To a large extent, this might not be considered a big surprise. 
The production
of photons of such high energies requires existence of accelerators of
parent charged particles with energy exceeding  
$10^{14} \, \rm eV$. Although the spectrum of 
the observed cosmic rays (CRs) 
extends to $10^{20} \, \rm eV$, it is quite possible that the efficiency
of acceleration of protons and nuclei in many  galactic sources,
for example SNRs,  
drops at $10^{14} \, \rm eV$~~\cite{Ces_Paris92}. 
Thus even in the case of sufficient target material, the $\pi^{0}$-decay
\gr emission above 10 TeV would be strongly suppressed. The same problem of
high-energy cutoff in acceleration spectra obviously exists also for
production of \grs due to the inverse Compton (IC) scattering of 
ultrarelativistic
electrons on the ambient low-frequency photons. 
In addition, the severe synchrotron  losses and reduction of the IC 
cross-section due to the Klein-Nishina effect makes this mechanism 
at such high energies significantly less efficient than
in the TeV energy region.   
From this point of view, nonthermal extragalactic sources like
powerful radiogalaxies which are able to accelerate 
particles 
well beyond $10^{15} \, \rm eV$~~\cite{Hillas_annrev84} might appear as more
promising objects (than the galactic sources) for $\geq 10 \, \rm TeV$
observations. However, the absorption of such energetic
\grs  in the diffuse  extragalactic background 
radiation (DEBRA) 
inevitably leads to strong attenuation  of fluxes of $\geq 10 \, \rm TeV$ 
\grs being produced beyond 100 Mpc (see Fig.~2). Thus any new (more 
effective) campaign of study of the sky in  $\geq 10 \, \rm TeV$ \gr photons
perhaps could be done only after significant  (by a factor of 10) 
improvement of   detector sensitivities compared to the sensitivities of 
current air shower arrays,  $J_\gamma \sim 10^{-14} \, \rm ph/cm^2 s$
at 100 TeV. Since the angular resolution of such arrays, as well as their 
ability to separate the $\gamma$-ray induced showers 
from hadronic showers produced by CRs are strongly limited, 
an essential improvement of
flux sensitivities of conventional air shower arrays can be achieved 
only by enlarging drastically the collection areas. This however  
seems to be not very attractive  approach.
More justified  perhaps could be a reduction
of the energy threshold of particle arrays 
towards 1 TeV by using water Cherenkov detectors like MILAGRO~\cite{milagro}.
Although the sensitivity of this detector is not very impressive (detection 
of the Crab requires $\sim$ 1 year  observation time) by the standards 
set by current imaging telescopes,
the large field-of-view and $100 \%$ duty cycle of MILAGRO would allow effective 
searches for TeV \grs from GRBs or for other serendipitous VHE phenomena. 
  
On the other hand,  the IACTs operating at large zenith angles may serve 
as an alternative method  for detection of $\geq 10 \, \rm TeV$ \grs
from persistent astrophysical objects.  
Indeed, this technique may provide, in addition to the good 
angular resolution and gamma/hadron  separation ability, 
huge detection areas~\cite{large_zenith}  exceeding the collection areas of 
the largest existing particle arrays by a factor of 10. 
The recent detection of \grs at energies up to $\geq 50 \, \rm TeV$
from the Crab Nebula (the first positive detection of the UHE 
gamma ray astronomy !) by  the CANGAROO  collaboration~\cite{cang_crab}
presents not only great astrophysical interest,   
but also demonstrates the high efficiency  of the technique.

\section{The Potential of Ground Based Gamma Ray Detectors}

\subsection{Single Imaging Telescopes}

The standard  IACT consists of a large (from a few to 10~m diameter) 
optical reflector equipped with  an array of  
photomultipliers in the focal plane of
the reflector, and fast (nanosecond) processing electronics. 
The field of view  of the camera should be at least $\sim 3^{\circ}$ in 
order to contain the images of showers with core position up to
150-200~m from the telescope. While the pixels with relatively modest size 
of about $0.25^{\circ}$ provide an adequate
quality of imaging of the air showers produced by TeV $\gamma$-rays,
at low energies ($E \leq 100 \, \rm GeV$), especially for telescopes
located at high mountain altitudes ($\geq 3$ km a.s.l.), the pixel
size should be close to $0.1^{\circ}$. A small pixel size is preferable
also for lowering (at the given area of the optical reflector) the energy
threshold, as well as for observations in the 
so-called regime 
of {\it large zenith angles}.  The operation of IACTs in this 
regime~\cite{large_zenith} can partially compensate the 
(typical) loss in \gr  statistics at 
high energies by  the significant increase of the  collection area  
when the source is observed at large zenith angles ($\geq 50^{\circ}$).      
Presently, two IACTs equipped with finest pixelization ($\sim 0.12^{\circ}$) 
imaging cameras, 
one in the northern (CAT)~\cite{CAT}, and another in the southern 
(CANGAROO)~\cite{cang_crab} 
hemispheres, can provide very effective study of \gr spectra of 
VHE sources beyond 10 TeV. 
%
%fig.3 (Mrk flares in May  1996)
\begin{figure}[t]
%\epsfxsize=13  cm
%\epsfysize=5.9 cm
%\epsffile[0 400 530 750]{fg3.ps}
\vspace{5.  cm}
\includegraphics{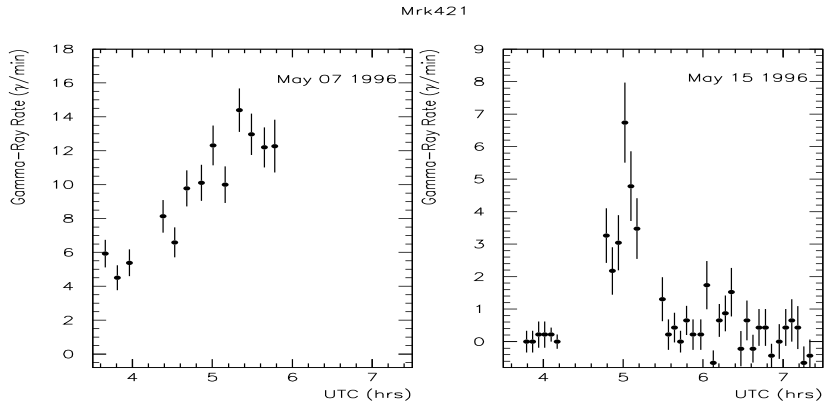}
\caption{Light curves of two Mrk~421 flare events of 1996 May 7 (left) and
1996 May 15 (right) [44]}
\end{figure}

A remarkable  feature of the atmospheric Cherenkov technique
is the large integration area of air showers, 
$A_{\rm eff} \geq 3 \times 10^{8} \, \rm cm^2$, a consequence of
the radius of the Cherenkov light pool $R_{\rm C} \geq 100 \, \rm m$. 
Since typically \gr fluxes rapidly increase  
at lower energies, the reduction of the energy threshold of Cherenkov 
telescopes would  yield a significant gain in the 
detected \gr statistics. 
For comparison,  the detection area of the GLAST~\cite{glast}
at GeV energies is  only $\sim 0.8 \, \rm m^2$. 
This implies that if the spectrum of a \gr source extends 
to the VHE region, the rate of detection of \grs by a single 
100 GeV threshold Cherenkov telescope would  exceed the GLAST detection rate 
even for very steep spectra of \grs with a differential power-law 
index $\alpha \leq 3.3$. Since the spectra of many potential 
\gr sources are expected to be 
significantly  flatter, this provides a principal basis for 
an astronomy {\it rich by \gr photon statistics}. 

However, this goal 
cannot be achieved without 
effective methods of suppression of the heavy background produced
by CR showers. Fortunately, the IACT technique does 
provide an adequate background rejection power~\cite{Hillas_HD,Fegan_97}.
Indeed, a typical  single imaging telescope can identify the  
electromagnetic showers produced by \grs from point sources
with efficiency  better than $99.7 \%$.
In particular, the   Whipple 10 m imager, today's  most sensitive 
single telescope,  provides 
$7 \sigma$ detection of the  Crab within only 1 h observation of the source. 
This implies that 100 h observations by this instrument 
could reveal point-like \gr sources above 250 GeV 
at the flux level of 0.07 Crab. 
This corresponds to 
the {\it energy flux}  $f_{\rm E} \sim 4 \times 10^{-12} \, \rm erg/cm^2 s$, 
which sounds impressive even by standards of the 
satellite-based \ga  (see Fig.~1).
Significant improvement of the performance of 
this telescope is expected in 1998. After the replacement of the 
old  camera by  a new,  very high resolution 541 pixel 
camera, the energy threshold will be lowered to 
$\sim 100 \, \rm GeV$, and  the flux sensitivity will be improved by a 
factor of three~\cite{Granite_Lamb}.

The `7-sigma-per-1 hour' signal from the Crab,
with the detection rate exceeding $100 \, \gamma$/h,    implies 
that any short outburst of a TeV source at the level exceeding  
the Crab flux can be discovered on timescales less than 30 minutes. 
This important feature of the IACT technique was convincingly demonstrated
by the Whipple group when two dramatic flares from Mrk 421 
were detected~\cite{W_Mkn421flare} in May 1996 (Fig.~3). 
During the first flare on May 7, lasted 
more than 2 hours, the \gr rate increased by a factor of 3, and 
reached the peak ten times the intensity of the
Crab. The second flare on May 15 was less intensive, but 
remarkable for its brevity - it lasted only 30 minutes !  
Note that detection of an energy flux comparable with the flux of 
May 15th flare 
($\sim 2 \times 10^{-10} \, \rm erg/cm^2 s$), but at GeV energies  by EGRET  
would require  a duration of the \gr outburst of about several days.

Further qualitative  improvement of the IACT technique
in the next few years is likely to proceed in two general 
directions: (a) implementation of the so called 
{\it stereoscopic approach}, and (b) reduction of the 
energy threshold of detectors towards the sub-100 GeV domain.

\subsection{Stereo Imaging}

The concept of stereo imaging is based on the simultaneous detection
of air showers in different projections by $\geq 2$ telescopes
separated by the distance comparable with the radius of the
Cherenkov light pool,  $R_{\rm C} \sim 100 \, \rm m$.  
Compared to {\it single} telescopes, 
the stereoscopic approach allows unambiguous and precise reconstruction
of the air shower parameters on the {\it event-by-event} basis.
In particular the accuracy of reconstruction of the direction 
of {\it individual}
\gr primaries can be as good as {\it few arcminutes}, which in the 
case of point-like
sources already provides suppression of the CR 
background by a factor
of 300 or more. A {\it comparable} reduction 
of CR contamination can be gained by exploiting  the 
{\it intrinsic differences} between the
electromagnetic and hadronic showers. Even though the images in different 
projections are not entirely independent of each other, the correlation is
only partial. Thus, in the case of point \gr sources 
the CR background could be suppressed by a factor of $\geq 10^{5}$.
In addition, the stereo observations provide effective suppression of
the background light of different origins (night sky backgrounds, local muons,
{\it etc.}), that ultimately results in significant reduction of the energy
threshold.  
%
%f4ab (HEGRA: angular characteristics)
\begin{figure*}[t]
\vspace{6. cm}
\includegraphics{fg4a.ps}
\includegraphics{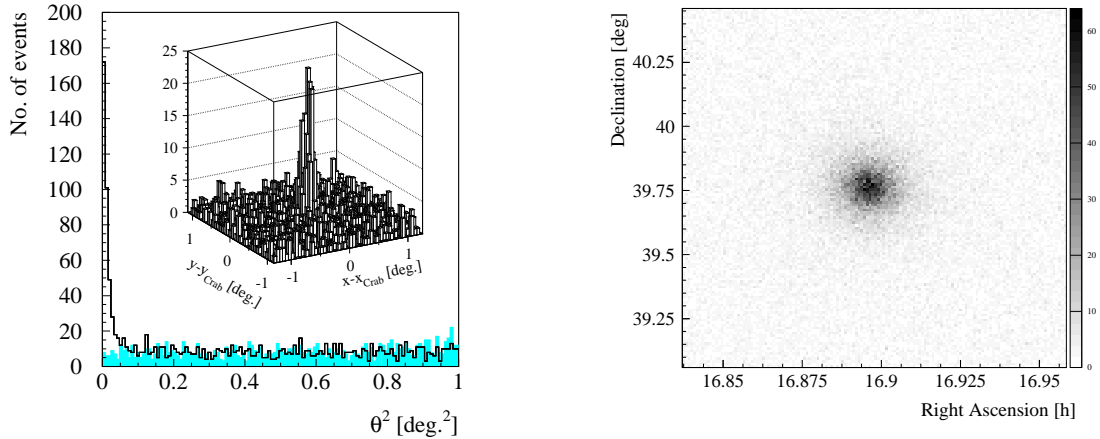}
\caption[]{The potential of the HEGRA IACT array for 
reconstruction of the arrival direction of primary $\gamma$-rays.
~{\bf a (left)}: Distribution of reconstructed directions of showers
detected in the 2-telescope  coincidence mode for 11.7~h 
ON-source observations of the Crab Nebula [45].
{\bf b (right)}: Two-dimensional plot of angular distribution of \gr emission
from the region around Mrk~501, after application of `shape' cuts.
The position of the maximum is consistent with the nominal position
of Mrk~501 within $0.01^{\circ}$.} 
\end{figure*}

The recent observations of the Crab Nebula and Mkr~501 
by the HEGRA stereoscopic system of 4 imaging telescopes 
located on Canary Island La Palma thoroughly confirm the early predictions
for the performance of the instrument. Fig.~4a shows the angular
distribution of showers detected from the direction of the Crab,
and selected after the image `shape' cuts~\cite{Daum_Crab}. 
The confinement of the \gr signal
from a point source in a very small region (with angular radius 
$\sim 0.1^{\circ}$) of the available two-dimensional phase space, 
in addition 
to the significant suppression of hadronic showers at the {\it trigger}
level, results in a strong $4 \sigma$-per-1~h signal already before the
gamma/hadron separation. The `shape' cuts provide further, by a factor of 100,
suppression of the hadronic background, while maintaining the efficiency
of the acceptance of \gr events at the level of $50 \%$. This leads to only 
$\sim 1$ background event in the `signal' region while the rate of \grs from
the Crab exceeds 20 events/1 h. This enables (1) an effective search for 
VHE \gr point sources with fluxes down 0.1~Crab at almost 
{\it background free} conditions, and thus drastic 
reduction of the observation
time (by a factor of 10)  compared to four independently operating 
telescopes, 
(2) detailed spectroscopy of strong (Crab-like) VHE emitters, (3)
study of the spatial distribution of \gr production regions on arcminute 
scales, (4) effective search for extended sources 
with angular size up to $\sim 1^{\circ}$ at the flux level of 0.1~Crab.  
The exploitation of the 
`nominal' 5-IACT HEGRA system with improved trigger condition 
should allow $5 \sigma$ detection of faint \gr sources 
at the flux level of 0.025~Crab which corresponds to 
$\sim 10^{-12} \, \rm erg/cm^2 s$ energy flux above the effective 
energy threshold of the instrument of about 
500 GeV~\cite{hegra_perf} (see Fig.~1).

The potential of the HEGRA stereoscopic system has been 
nicely demonstrated  by the measurements of the flux, spectrum, 
and variability
of TeV \grs from Mrk~501 during the state of high activity of the source 
started in March 1997.
The almost background free detection of \grs with a rate exceeding
$100 \, \gamma / \rm h$, coupled with energy resolution of the instrument
$\sim 20 \%$, allowed a study of the flux variation
on time scales between 5 min to days~\cite{hegra_mkn501_AA} 
(Fig.~5), evaluation of
the differential energy spectrum based, during the strong flares,  
only on few hours of  observations~\cite{hegra_perf} (Fig.~6),
and location of the position of the source within the statistical error of
30 arcsec (Fig.~4b).

%f5 (HEGRA: Mkr 501/variability)
\begin{figure}[h]
\vspace{5. cm}
\includegraphics{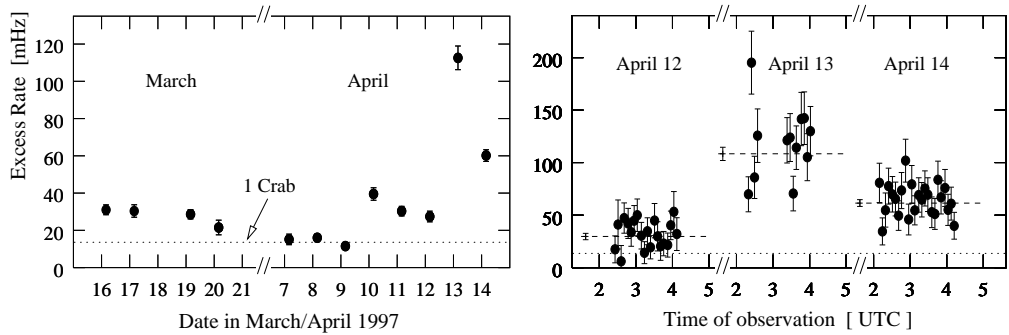}
\caption[]{Detection 
rate of \grs from Mkr~501 by the HEGRA IACT system on 
a night-by-night basis,
for the whole data set (left), and in 5 minute intervals for the 
last three nights (right) [5].}
\end{figure}
%
%f6 (HEGRA: Mkr 501/spectrum)
\begin{figure}[h]
\vspace{5. cm}
\includegraphics{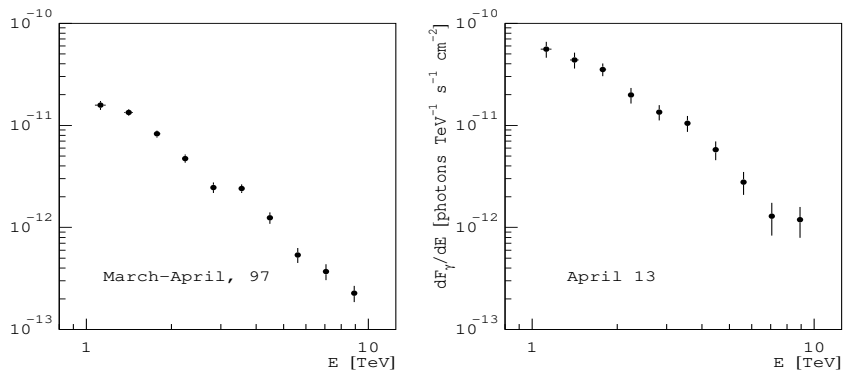}
\caption[]{Differential energy
spectra of Mkr~501 measured by the HEGRA IACT system [46] 
during the strong flare on April 13, and  averaged 
over the whole March-April period shown in Fig.5.}
\end{figure}

The continuous monitoring of Mkr~501 from March
until September 1997 by the CAT, HEGRA, and Telescope Array groups
showed that the source was in very active state with
variable flux exceeding in average the Crab flux 
by a factor of 2 or more.
The long-term behavior obtained  by the Telescope Array 
is presented in Fig.~7. It shows not only
dramatic day by day variability, but also a possible $\sim 25$ day
(quasi) periodicity in the signal~\cite{TA_mkn501}.

More than 30,000 (!) \grs  up to 10 TeV and beyond  have been detected by the
HEGRA IACT system during several months of observations. This 
demonstrates the remarkable feature of the IACT technique as a tool
providing an {\it astronomy rich by photon statistics}.

\subsection{Future 100~GeV Class IACT Arrays}

One of the important issues for future detectors is the choice of the energy
domain based on two principal arguments: (a) the {\it astrophysical 
significance} (goals) and  (b) the {\it experimental 
feasibility/reliability} (cost). If one limits the energy region 
to a relatively modest threshold around 100 GeV, the performance 
of IACT arrays and their practical implementation 
can be prediced with high confidence. In practice, an energy threshold of 100 GeV
can be achieved by a stereoscopic system of IACTs consisting of 
optical reflectors with diameter $\sim 10 \, \rm m$ like the
Whipple mirror~\cite{Whipple_dish}, 
and equipped with conventional PMT-based high resolution cameras 
like the cameras recently built 
by the CAT~\cite{CAT_camera} and HEGRA~\cite{HEGRA_camera} groups.
%  
%f7 (TA-mkn501 light curve)
\begin{figure}[t]
\vspace{3.7cm}
\includegraphics{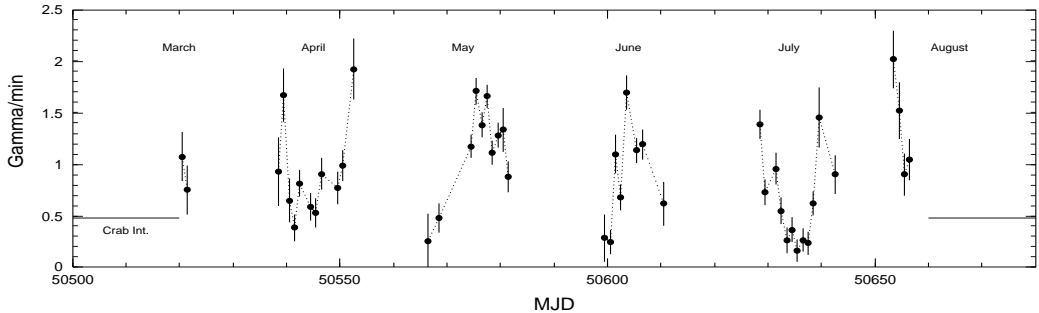}
\caption[]{The light curve of Mrk~501 obtained by the Telescope Array [47].}
\end{figure}
After reaching the maximum possible suppression of the CR background by
simultaneous detection of air showers in different projections -- limited
basically by intrinsic fluctuations in cascade development -- the further
improvement of the flux sensitivities for a given energy threshold 
can be achieved by the increase of \gr statistics, i.e. by increasing the 
shower collection area. For  a single telescope the effective 
detection area is determined essentially by the radius 
of the Cherenkov light pool. 
However, the effective  shower detection area of the multi-telescope systems 
is determined by the total geometrical area of the array
which 
may be gathered 
from individually triggered groups of telescopes, IACT {\it cells},  
with an optimum linear size of the cell $L \sim R_{\rm C} \sim 100 \, \rm m$. 
The design of the {\it cell} depends on the specific
detection requirements. However, since the quality of reconstructed shower 
parameters continues to be improved noticeably up to three or four 
telescopes operating in coincidence~\cite{Array}, the 
optimum design of the {\it cell} 
seems to be a triangular or quadrangular 
arrangement of IACTs. The stereoscopic 
approach with its powerful suppression of CR background 
requires  cameras with relatively modest, e.g. $0.25^{\circ}$
pixel size, and thus allows allocation of the available channels 
for enlarging the FoV to $\geq 5^{\circ}$, in particular for effective 
sky surveys, as well as for the increase of  
the collection area up to $1 \, \rm km^2$. 
A possible design of multi-cell array 
of 100~GeV class IACTs is demonstrated  in Fig.~8a. The range of  
expected energy flux sensitivities studied recently by comprehensive 
Monte Carlo simulations~\cite{Array} is shown in Fig.~1. 

Perhaps the idea of `$1 \, \rm km^2$' array of telescopes 
sounds too ambitious,  but it is not unrealistic. In fact, such an array, 
containing as many as 100 telescopes
each with an aperture 3~m and a 1024 pixel camera, has been proposed
already in 1992 by the Tokyo group~\cite{TArray_1992}, 
with a dual aim of (1) detecting 
highest energy CRs above $10^{18} \, \rm eV$ and (2) observing
TeV \gr sources. Although  the successful detection 
of the Crab Nebula and Mrk 501 by the prototype 5-telescope-array 
in Dugway (Utah) is very  encouraging~\cite{TA_mkn501}, 
for future purposes of  {\it \ga} it perhaps would be important  
to reduce the energy threshold of the telescopes, e.g. by 
using larger optical reflectors.

Among a variety of competing projects,  the design of 
the 100 GeV class IACT array is the most likely to 
obtain a qualitative improvement in performance at an
affordable cost~\cite{AhAk_annrev97}.
Presently two projects of similar philosophy,  VERITAS 
({\bf V}ery {\bf E}nergetic {\bf R}adiation 
{\bf I}maging {\bf T}elescope {\bf A}rray {\bf S}ystem)~\cite{veritas} 
array of nine 10m diameter telescopes in the USA (Arizona),  
and HESS ({\bf H}igh {\bf E}nergy {\bf S}tereoscopic 
{\bf S}ystem)~\cite{hess} 
array of sixteen 10m telescopes to be installed in Spain or Namibia, 
are intensively discussed
as two major instruments of the ground-based \ga at the  entrance into  
the next millennium.  

\subsection{Sub-100 GeV Ground-Based Detectors}

The strong scientific motivations to fill the gap between 
the space-based and ground based observations
recently stimulated several interesting proposals for extending the
atmospheric Cherenkov technique to sub-100 GeV 
region~\cite{Lamb_Snowmass}.  Perhaps this ambitious goal 
could be best addressed by the same (above described)  
stereoscopic systems of IACTs but with  the telescopes aperture 
$S_{\rm ph.e.}=S_{\rm mir} \chi_{\rm ph \rightarrow e} \geq 50 \, \rm m^2$,
where $S_{\rm mir}$ is the mirror area and 
$\chi_{\rm ph \rightarrow e}$ is the photon-to-electron conversion
factor (quantum efficiency) of the optical detectors.  
With development of the technology of novel, fast optical 
radiation detectors of high quantum efficiency 
($\chi_{\rm ph \rightarrow e} \geq 50 \%$) and   
design of (relatively inexpensive) 20 m class reflectors with 
adequate optical quality for the Cherenkov imaging on scales 
of several arcminutes,  it would be possible to
reduce the detection threshold down to 20 GeV,  or even 10 GeV
for an IACT system installed at high mountain 
elevations (e.g. 3.5 km a.s.l).
Considerable R\&D  efforts, especially by the 
Munich group \cite{Magic}, are already started
in both directions. If successful, this
activity  will result  
in construction of a large single reflector 
telescope (MAGIC) \cite{Magic} which, together with two ongoing 
projects of low  energy threshold Cherenkov telescopes based on
large mirror assemblies of the existing solar power plants in 
the USA (STACEE \cite{STACEE}) and France (CELESTE \cite{CELESTE}),
will start to explore the energy region 
between 20 and 100 GeV. 
Moreover, the MAGIC telescope could be considered 
as a prototype for the basic element in  
future (post `VERITAS/HESS')  arrays  of 10 GeV class IACTs.
Few MAGIC-type  telescopes being combined in the 
stereoscopic system could not only intervene the domain of  
satellite-based gamma ray astronomy, but also could provide very large 
detection areas at higher energies,  especially when observing 
\gr sources at large zenith angles. If high efficiency of this
technique,  that already has been 
demonstrated by the CANGAROO group at multi-TeV energies,   
could be extrapolated to the 0.1-1 TeV region, the use of an
array consisting of few  10~GeV class IACT systems 
(with 3 or 4 telescopes in each) could cover very broad energy 
range extending  from 10 GeV to 10 TeV.

A possible arrangement, operation modes,  and predicted performance  
of an array of 10 GeV class telescopes  
are described  in Fig.~8b.
The expected range of the energy flux sensitivities of 
such arrays is shown in Fig.~1.

%\newpage
%
%fig.8 (future arrays)
\begin{figure}[t]
\vspace{13. cm}
\includegraphics{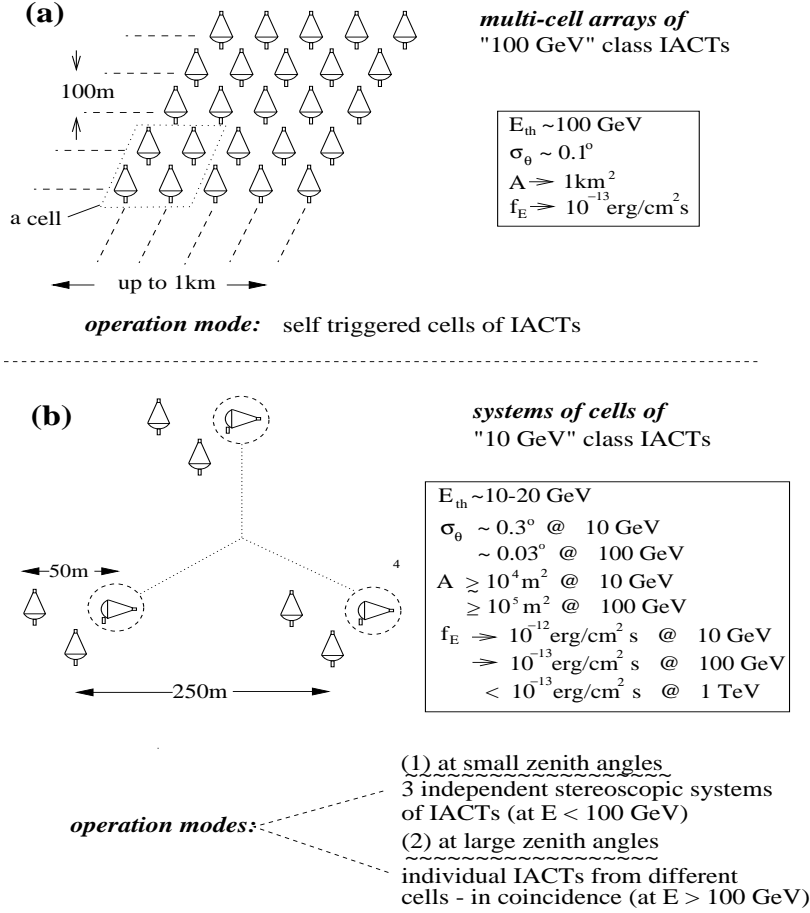}
\caption{Two versions of future stereoscopic IACT arrays}
\end{figure}

\section{VHE Gamma Rays -- Carriers of Unique Astrophysical Information}

With future systems of atmospheric Cherenkov telescopes, ground-based \ga
will enter a new era with an ambitious aim of deep
study of the \gr sky in broad energy region that extends from 10 GeV
to $\geq 10 \, \rm TeV$. In particular,

$\bullet$~Very large arrays of stereoscopic IACTs
will be able to probe, typically during several weeks of 
observations, the potential VHE emitters of \grs at an 
{\it energy flux}~\footnote{
The energy flux characterizes  the relative release
of the power (luminosity) of a source at different wavelengths
(for an isotropically emitting source at a distance $d$, 
$L=4 \pi d^2 f_{\rm E}$).
Therefore $f_{\rm E}$ is a meaningful and objective parameter, 
when comparing the sensitivities of instruments 
operating in different energy domains.}
level
$f_{\rm E}=E^2 dJ/dE \sim 10^{-13} \, \rm erg/cm^2 s$, i.e.
by about three orders of magnitude lower than the
sensitivities  achieved at MeV and GeV energies 
by the current satellite-borne 
instruments. 
This allows study of different VHE source populations 
with apparent ($4 \pi$) luminosities  
$L_{\rm VHE} \geq 10^{37}(d/ \, \rm 1 Mpc)^2 \, \rm erg/s$ -- 
a very impressive achievement  even by the 
standards of the classical branches of astronomy.

$\bullet$~The high sensitivity of multi-IACT arrays
partially compensates the inefficiency of the technique 
regarding all-sky surveys~\footnote{Note, however, that 
presently some interesting ideas are 
circulating about the design of
an imaging `All Sky Monitor' telescope at VHE energies with FoV
of about one steradian (T.Kifune, private communication).}.  
In particular, during a 1-year survey it will be possible to cover
more than 1 steradian of the sky
at the source detection threshold $\leq$ 0.1~Crab 
($f_{\rm E} \leq 10^{-11} \, \rm erg/cm^2 s$). 

$\bullet$~The multi-IACT arrays are very effective instruments 
for the search for highly sporadic nonthermal phenomena from different 
celestial objects
which may appear below the detection threshold of 
satellite-borne instruments at MeV and GeV energies. 
In particular, the minimum
detectable {\it fluence} $\leq 10^{-8} \, \rm erg/cm^2$ 
of a burst lasting $\leq 1 \, \rm h$ would allow detection of possible
VHE counterparts of GRBs if even only a small part of the total
energy of the event is released in $\geq 100 \, \rm GeV$ photons.

$\bullet$~The IACT arrays
provide the best angular resolution in
gamma-ray astronomy from MeV to ultra-high energies, 
$\delta \theta \leq 0.1^{\circ}$, and reasonable
spectroscopic measurements with energy resolution $\Delta E/E \leq 20 \%$.  

\vspace{1mm}

The expected performance of detectors
as well as the recent theoretical/phenomenological predictions supply a 
strong rationale for the systematic study of primary $\gamma$-radiation
in the VHE domain. Such efforts will provide crucial insight into the 
understanding of ultrarelativistic processes in astrophysical
settings, on both galactic and extragalactic scales.

\subsection{Origin of Galactic Cosmic Rays}

{\sc shell-type SNRs}.~Historically the first interest 
in \ga arose in connection with 
general idea on the crucial role of \gr observations in solving the 
problem of origin of cosmic rays (see e.g. [59, 60] and 
references therein). The future IACT arrays, with their excellent
flux sensitivity and angular resolution, should finally address this 
fundamental issue. 
Of primary concern is the decisive test~\cite{DAV} of the hypothesis that the
SNRs are responsible for the bulk of observed CRs up to 
$10^{14} \, \rm eV$.
Conservative phenomenological estimates~\cite{DAV,Naito}, as well
as recent detailed calculations~\cite{Berezhko} show that 
in the framework of diffusive shock acceleration many 
shell-type SNRs could be visible in VHE photons. 
Since the spectrum of the shock-accelerated 
particles is believed to be hard, 
${\rm d}N/{\rm d}E \propto E^{-\alpha}$ with $\alpha \sim 2$,  
the flux of VHE $\gamma$-rays, produced  in a SNR at a distance $d$ 
due to interactions of accelerated protons with the ambient gas of density $n$,
is easily estimated as  
$J(> E) \approx 10^{-11} \, \rm (E/1 \, {\rm TeV})^{-1} \,
(W_{\rm CR}/10^{50} \, {\rm erg})(n/1 \, {\rm cm^{-3})}
(d/1 \, {\rm kpc})^{-2} \, \rm ph/cm^2 s$,
where $W_{\rm CR}$ is the total energy in accelerated CRs.
The peak luminosity of \grs typically appears at the so-called
late sweep-up phase~\cite{Berezhko}, 
i.e. $10{^3}-10^{4}$ years after the SN explosion, when the radius of SNRs 
exceeds several parsecs. This results in a typical angular size of 
relatively close ($d \leq 1 \, \rm kpc$) SNRs of about $1^{\circ}$.
Although the conflict  between the angular size 
($\phi \propto 1/d$) and the \gr flux ($J \propto d^{-2}$)
significantly limits the number of 
SNRs detectable in $\gamma$-rays, the flux sensitivities of 
future IACT arrays
are quite sufficient for detection of  VHE \gr signals from 
SNRs if $(n /1 \, {\rm cm^{-3})} (d/1 \, {\rm kpc})^{-2} 
\geq 0.1$ (see Fig.~9a).

%
%fig9ab (snr and diffuse)
\begin{figure*}[t]
\vspace{5.4 cm}
\includegraphics{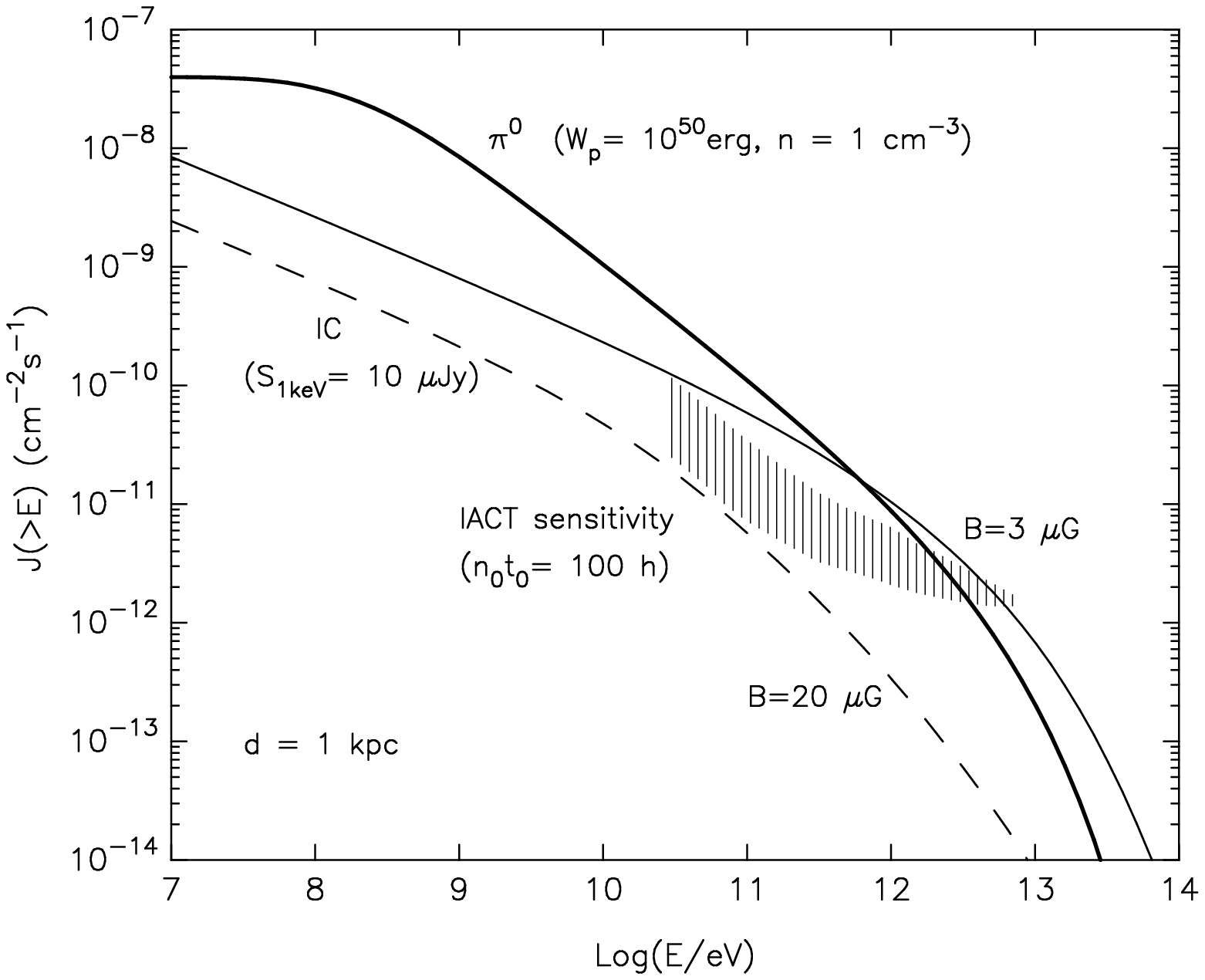}
\includegraphics{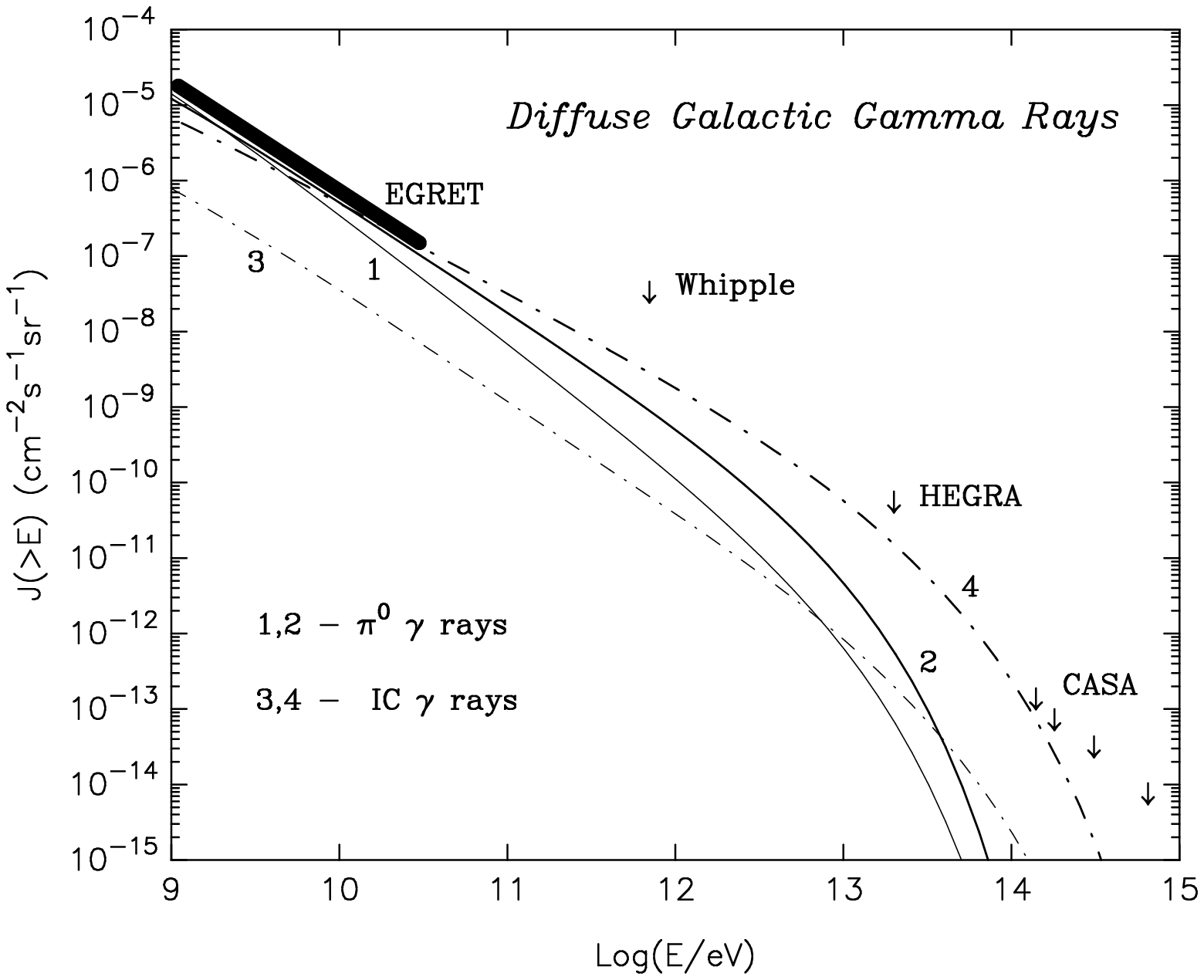}
\caption[]{$\pi^0$-decay and inverse Compton \gr fluxes.
{\bf a (left):}~ The predicted \gr fluxes from a 
SNR located at $d=1 \, \rm kpc$.
The $\pi^0$-decay \gr fluxes are normalized to the total energy in accelerated
protons $W_{\rm CR}=10^{50} \, \rm erg$, and density of the ambient gas 
$n=1 \, \rm cm^{-3}$. The IC \gr fluxes are normalized to the synchrotron X-ray 
flux at 1 keV $S_{\rm 1 keV}=10 \, \rm \mu Jy$ 
for ambient magnetic fields $3 \, \mu \rm G$
and $20 \, \rm \mu G$. Both the electron and proton acceleration spectra are 
taken in the form $\propto E^{-2} \exp(-E/100 \, \rm TeV)$.
The hatched region shows the IACT flux sensitivity corresponding to sources 
with angular size~ $\leq 0.1^{\circ}$~ (lower boundary) to 
$\sim 1^{\circ}$ (upper boundary) during observation time~$t_0$~   
by an array consisting of 
$n_0$ IACT cells; $n_0 \cdot t_0=100 \, \rm h$. {\bf b (right):}
Diffuse fluxes of  \grs from the direction of the galactic center 
for galactic latitudes $|b| < 10^{\circ}$.} 
\end{figure*}

When extracting information about the shock accelerated protons,
one has to subtract a possible
non-negligible contribution to the $\gamma$-ray fluxes
from  directly accelerated electrons which upscatter  the photons of
2.7~K background radiation to \gr energies.
Typically the contribution of IC \grs
is below of the fluxes of $\pi^{0}$-decay photons (except
for the SNRs with low gas density, $n \leq 1 \, \, \rm cm^{-3}$
and/or very low magnetic field, $B \sim B_{\rm ISM} \sim 5 \mu \rm G$)
if one takes into account that the
nonthermal synchrotron X-ray fluxes from most of shell type SNRs
do not exceed $S_{\rm 1keV}=10 \, \mu \rm Jy$ ($f_{\rm x} \approx 
2.5 \times 10^{-11} \, \rm erg/cm^2 s$) (see Fig.~9a).

Recent discovery of nonthermal  
X-rays from SN1006~\cite{ASCA_1006}, which (almost unambiguously) 
implies the existence  of electrons of energies  $\geq 10 \, \rm  TeV$, 
motivated optimistic predictions about 
detectable VHE \gr fluxes of IC origin 
from  this shell-type SNR~\cite{masti_1006}.
Indeed,  this object  recently 
has been claimed to be seen in TeV \grs \cite{cang_sn1006}, the estimated \gr 
flux being in a reasonable agreement with the theoretical calculations.
Nevertheless, it is perhaps premature to draw final conclusion on 
the IC origin of the observed TeV radiation.
In order to match the synchrotron X-ray and IC \gr fluxes,
the magnetic field in the shell should not exceed $10 \mu \rm G$.
This is a rather bold requirement, being, for example,    
by a factor of 10
less than the value estimated from the condition of equipartition 
between the  magnetic field and relativistic electron pressures. Obviously,
the IC models of $\gamma$-radiation of SN1006
need independent arguments 
in favor of such small B-field in this young SNR. 
Otherwise, the remaining possibility could be 
$\pi^{0}$-decay origin (?!) of detected TeV radiation.      

Detection of $\pi^0$-decay  \grs from 
SNRs would be the first straightforward proof of the 
shock acceleration of CR protons and nuclei in 
these objects. On the other hand, a 
failure to detect $\pi^0$-decay \gr signals from several selected 
SNRs would impose constraint on the energy in accelerated protons
$W_{\rm CR} \leq 10^{49} \, \rm erg$. Contrary to current belief, 
this would indicate the inability of the ensemble of galactic SNRs to explain
the observed CR fluxes.  

Finally, another interesting possibility should be mentioned in 
the general context of visibility of SNRs in VHE $\gamma$-rays.  
In the case of explosion of a massive star as a supernova, 
the \gr luminosity could be extremely high during several months or more  
after the explosion~\cite{Ber_Ptus,KDB_youngsnr} 
due to (1) very effective particle acceleration
at the shock front and (2) existence of relatively dense 
confining medium. 
In particular, \gr fluxes as high as 
$10^{-12} \, \rm ph/cm^2 s$
above 1~TeV have been predicted~\cite{KDB_youngsnr} 
for very powerful radio supernova 
SN1993J situated in the galaxy  M81 at a distance 
$d \approx  3.6 \, \rm Mpc$. 
If so,  the future IACT arrays with their potential to find
faint \gr sources with luminosities 
as low as $L_{\rm VHE} \sim 10^{37} \, (d/1 \, \rm Mpc)^2 \, \rm erg/s$,
should be able to detect positive VHE \gr signals from recent extragalactic 
SNRs within the local group of galaxies at the early stages 
($\leq 1 \, \rm yr$) of their explosion.

\vspace{2mm}
\noindent
{\sc GMCs as tracers of galactic cosmic ray accelerators.}~SNRs 
are still only 
one of the possible CR accelerators,   
and it is quite possible that different other classes of CR 
sources contribute comparable fractions to the observed CR flux.
Thus, only direct identification of these sources as particle accelerators
through their characteristic \gr emission can help to elucidate the origin
of galactic CRs. The existence of an accelerator is by 
itself not enough for effective \gr production; one needs the 
second component, a {\it target}. The so-called giant molecular clouds (GMCs),
with $10^{4}$ to $10^{6}$ solar masses, seem to be ideal objects
to play that role. These objects are genetically connected 
with active star formation regions --  the most probable settings
(with or without SNRs) of CR production. At certain epochs, 
depending on the distance from the 
accelerator with total CR energy   
$W_{\rm CR} \geq 10^{49} \, \rm erg$, many clouds may become detectable in VHE 
$\gamma$-rays~\cite{ahat_clouds}. 
Therefore, surveys of the galactic plane, that search
for GMCs irradiated by CRs arriving from young, nearby 
accelerators may provide unique information about CR sources 
and their distribution in the Galaxy. Another interesting issue 
would be detection of diffuse $\pi^0$-decay $\gamma$-radiation 
of the galactic disk formed due to superposition 
of contributions of faint (individually invisible) GMCs.   
  
\vspace{2mm}

\noindent
{\sc diffuse galactic gamma ray background.}
The EGRET measurements of the diffuse background in the
direction of the inner Galaxy show an excess in the GeV \gr flux,
compared with the predictions (curve 1 in Fig.~9b)
based on the assumption that the average
spectrum of CRs in the Galaxy is represented by the locally measured
CR spectrum~\cite{egret_galdif}. 
This effect  formally can be explained by assuming harder 
(e.g. $\propto E^{-2.5}$) proton spectra in the regions where 
the bulk production of \grs takes place (curve 2 in Fig.~9b). 
Actually such
assumption could be well justified by speculating that 
the main part of the observed diffuse \gr background is produced 
selectively, i.e. it is a result of radiation  originating in 
regions which contain particle accelerators and 
massive gas clouds.
Indeed, if \grs are produced at the interaction of a cloud with 
relatively fresh (recently accelerated) particles with hard spectra 
which have not yet suffered strong modulation (steepening) due to
the propagation effects of CRs in the interstellar medium, the resulting
\gr spectra could be significantly harder that the typical \gr 
spectra produced by the `sea' of galactic CRs~\cite{ahat_clouds}.
 
The excess above a few GeV in the diffuse galactic spectrum
can be explained also by the IC photons. Note that this hypothesis requires
very strong (by a factor of 10) enhancement of the average flux of 
$\geq 100 \, \rm GeV$ electrons in the galactic disk  
compared with the measured flux of VHE 
electrons~\cite{electrons_nish}. However,  
formally it cannot be ruled out if one takes into account that the 
directly detected electrons of such high energies are produced
within the  nearest 100~pc region, and therefore do not 
carry information about the electron fluxes on the galactic (kpc) scales.
This possibility is 
demonstrated  in Fig.~9b where the IC \gr flux calculated
for the locally observed electron spectrum (curve 3) 
is shown together with the IC flux calculated for some hypothetical
spectrum of galactic electrons (curve 4) assumed in order to
match the  measured \gr fluxes at GeV energies.

While the predictions of diffuse galactic fluxes  
are still below the upper limits set both at TeV and PeV energies, 
the new stereoscopic IACT systems  should be able, however,
to probe the 
predicted range of diffuse $\pi^0$-decay and IC \gr 
fluxes shown in Fig.~9b.

\vspace{2mm}

\noindent
{\sc pulsars and plerions.}~ Presently six radiopulsars
are firmly established  by
EGRET as GeV \gr emitters~\cite{EGRT_puls}.
Modulation of  \gr light curves at the periods
of rotation of these objects indicates
that $\gamma$-rays are produced near the surface of neutron stars
due to  nonthermal cascade processes supported by
curvature radiation, inverse Compton scattering
and synchrotron radiation of accelerated 
electrons either at the polar cap or in the vacuum gaps
at the outer magnetosphere (for review see e.g. [72]). 
Irrespective of details of
the models, it is likely that flat GeV \gr spectra
should significantly steepen at higher energies
due to  attenuation caused by pair production in the pulsar
magnetic field.  
The position of the spectral turnover depends
essentially on the localization of the \gr production region(s).
Thus, the spectrometric measurements at 
energies above 10~GeV  may provide a crucial test
for different scenarios of particle acceleration 
in the pulsar magnetospheres.
Since the fluxes of the EGRET pulsars at several GeV
exceed  $10^{-8} \, \rm ph/cm^2 s$,  the future low-threshold ground based detectors,
with predicted flux sensitivity  $\sim 10^{-10} \, \rm ph/cm^2 s$
at energies between 10 - 30~GeV, will be able to provide such important
information. 
%
%fig10 (pulsar)
\begin{figure}[t]
\vspace{5.5 cm}
\includegraphics{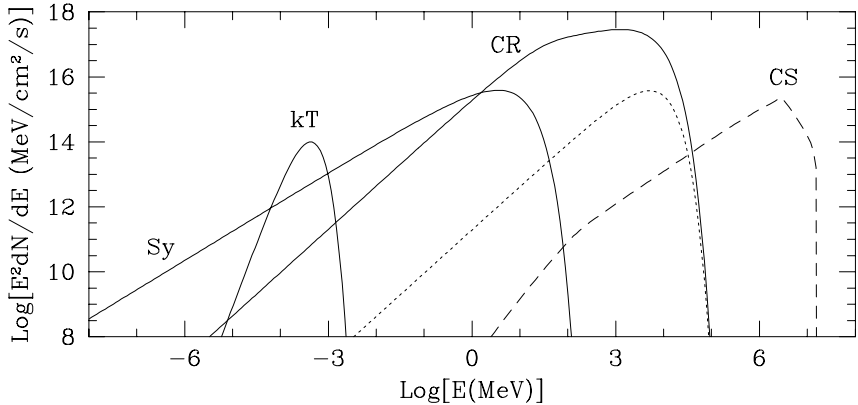}
\caption[]{Characteristic phase-averaged spectrum for a young \gr
pulsar. Solid lines show the curvature (CR), synchrotron (Sy), and the thermal
surface flux (kT). The dashed curve (CS) gives the TeV pulsed spectrum from
Compton upscattering of the synchrotron spectrum on the primary $e^{\pm}$ 
[73].}
\end{figure}

The search for TeV \grs from pulsars
was started many years ago, 
but despite several claims of possible detections
of pulsed TeV radiation, in particular from the Crab and Geminga pulsars,
observations with more sensitive imaging telescopes 
failed to confirm these early results. 
This is in agreement with the models
which predict cutoffs of the  radiation  
beyond 10~GeV (see Fig.~10). Meanwhile the recent developments
of pulsar models predict the  existence of
a new component of $\gamma$-radiation produced due to 
the inverse Compton mechanism in the outer magnetosphere~\cite{Romani}.
An interesting feature of this radiation is its hard spectrum below 1 TeV, 
with  sharp cutoff above $\sim 3 \, \rm TeV$. Thus,  the most 
promising energy region
for detection of this component seems to be a 
rather narrow interval around 1 TeV.
The pulsed TeV flux contains somewhat less 
than $1 \%$ of the pulsed GeV flux,
and therefore some pulsars (like Vela and Geminga) 
could be considered as promising
targets for future air Cherenkov experiments~\cite{Romani}. 
 
Besides the pulsed radiation produced in the pulsar magnetospheres, 
one may expect \grs from more extended regions - pulsar 
driven nebulae (plerions).
Relativistic electrons accelerated by the pulsar 
wind termination shock~\cite{Arons}
up to energies $\geq 10 \, \rm TeV$ interacting 
with nebular magnetic and photon
fields produce bright synchrotron and IC \gr emission.
The Crab Nebula is the most powerful 
representative of this source population
with nonthermal emission observed over 21 decades 
of frequency from 
$10^{7}$ to $10^{28}$~Hz. Due to the large 
magnetic field of the Crab Nebula, 
$B \geq 10^{-4} \, \rm G$, only $\sim 0.1 \%$ 
of the energy of electrons is converted
to IC \grs, the rest being radiated in the form 
of radio to X-ray synchrotron photons~\cite{atah_crab/mnras}.
The low efficiency of \gr emission of the Crab 
is compensated by the extremely high
rate injection rate ($\sim 5 \cdot 10^{38} \, \rm erg/s$) of 
relativistic electrons by the pulsar into the nebula.  This is the only 
reason why we see the Crab Nebula as bright 
TeV \gr emitter. The magnetic fields
in other plerions are likely to be one or two 
orders of magnitude smaller, which 
makes these objects more effective \gr 
emitters, since the radiative energy loss
of electrons is shared almost equally between 
the synchrotron and IC channels ($f_\gamma/f_{\rm x}=w_{\rm MBR}/w_{\rm B}
\simeq 1 \, (B/3 \mu \rm G)^{-2}$).
Note that since the IC cooling time of electrons
$t_{\rm IC} \propto 1/E$, this process works 
most effectively in the VHE domain.
This raises the hope that, with forthcoming 100~GeV class IACT arrays 
with energy flux sensitivity
of about $10^{-13} \, \rm erg/cm^2 s$, IC $\gamma$-radiation will be
observed from many 
plerions~\cite{Harding_DH,ahatkif_mnras,SNRrev_GRO4} even though  the 
pulsar spin-down
luminosities of these objects does not exceed a few percent of the power
of the Crab pulsar. In this context, the synchrotron X-ray
nebulae recently discovered by the ASCA and ROSAT satellites 
around many radiopulsars~\cite{Xnebula_asca} 
at the flux level $f_{\rm x} \geq 10^{-12} \, \rm erg/cm^2 s$
seem to be very attractive targets for VHE \gr observations. 
Detection of unpulsed TeV radiation by the
CANGAROO group from the direction of 
Vela~\cite{cang_vela} and PSR~1706-44~\cite{cang_1706}
is likely to confirm this optimistic view. 

\subsection{Extragalactic Sources of VHE gamma Rays}

\noindent
{\sc active galactic nuclei.}~The discovery of $\geq 50$  \gr emitting AGNs by
EGRET demonstrates that high energy
$\gamma$-radiation above $100 \, \rm MeV$ is a 
common feature of blazars -- BL Lac objects, radio loud AGNs, and optically
violent variable (OVV) quasars~\cite{von_Mont}.
The huge apparent
\gr luminosities and  rapid variability of the EGRET AGNs
favor a strongly anisotropic character of
high energy radiation which may be naturally
attributed to the relativistic bulk motion of
the jets, 
ejected from the central compact source, and  
directed almost
along the observer's line of sight.
Also, the anisotropic character of $\gamma$-radiation
seems to be the only way to avoid
internal  photon-photon absorption, 
as well  as to reduce  the energy requirements to
the central engines to a level which could be explained by
accreting massive black holes --
most likely powerhouses of AGNs.
Even so, attenuation of $\gamma$-rays
inside of distant (and typically most luminous) AGNs with redshift $z \geq 1$
is unavoidable well above 10~GeV.
In addition,
VHE radiation above 100~GeV emitted from cosmologically
distant sources is absorbed in the intergalactic space.
This determines the importance of sub-100~GeV
ground-based detectors for the study of relativistic processes in distant AGNs. 
%
%fig11 (mrk501 whipple/osse/ASM)
\begin{wrapfigure}[26]{l}{8. cm}
\epsfxsize=7 cm
\epsffile[45 48 530 740]{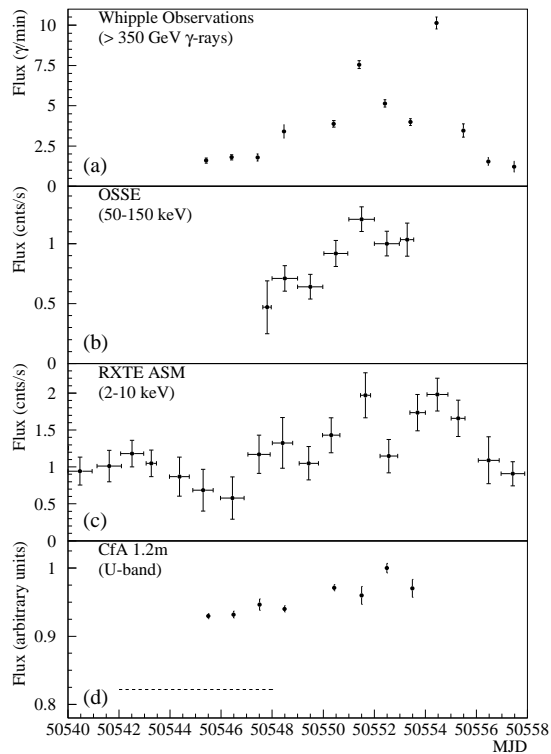}
\caption[]{TeV (a), hard X-ray (b), low energy X-ray (c),
and optical (d) light curves of Mrk~501 for the period 1997 
April 2 (MJD 50540) to April 20 (MJD 50558) [93]} 
\end{wrapfigure}

Many famous extragalactic sources representing different AGN 
populations, and located within $\approx 10^3 \, \rm Mpc$,
can be considered as first priority targets for 
observations above 100~GeV.
First of all this concerns the BL Lac objects~\cite{urry} 
two representatives of which, Mrk~421 and
Mrk~501, are already
established as TeV emitters. 
Different mechanisms have been suggested for \gr production 
in AGN jets, attributed to both 
{\it hadronic}~\cite{mannheim,dar_agn} and 
{\it electronic}~$^{83-89}$ interactions.  
%~\cite{derm_schi,sikora,marscher,maraschi,takahara,masti_kirk,bed_prot}. 
The variability of  Mrk~421 and Mrk~501  
on timescale $\sim 1$ day or less 
($\Delta t_{\rm obs} \sim$15 min (!) during the flare of Mrk~421 on May 15; 
see Fig.~3) support the
{\it electronic} models which assume that the \grs  are produced due to interactions 
of electrons, accelerated in the jet, with the ambient 
photon fields. For characteristic jet parameters only the IC mechanism
seems to be able to provide radiative cooling time shorter than the observed duration
of the flares (in the jet frame 
$\Delta t^{\prime}=\delta_{\rm j} \cdot \Delta t_{\rm obs}$). 
Meanwhile, a possibility for the {\it hadronic} models still  remains
within the scenario of ``moving target crosses beam'' (suggested originally
for X-ray binaries in much smaller (galactic) scales~\cite{ahat_beam/target}),
assuming that the episodic behavior of TeV radiation is due to the 
fast moving gas clouds
crossing the jet~\cite{dar_agn,bed_prot} 
and/or destruction (evaporation) of the targets under the
powerful jet.   

The discovery by ASCA and Whipple
of correlated flares of Mrk~421 in keV and TeV energy 
bands~\cite{mkr421_asca,mkr421_whipple}, as well as 
the recent multiwavelength observations of Mrk~501 in 
April 1997~\cite{mkr501_catanese,mrk501_pian} (see Figs.~11 and 12),  
indicate that both components originate in
a relativistic jet, with Doppler factor 
$\delta_{\rm j} \geq 10$, 
due to synchrotron and IC radiation of
the same population of 
electrons.  Remarkably, since the
IC cooling time
$t_{\rm C}=-E_{\rm e}/(dE_{\rm e}/dt) \propto E_{\rm e}^{-1}$, and
the Compton scattering boosts the ambient
photons of energy $\epsilon_0$ to 
$E_\gamma \approx \epsilon_0 \cdot (E_{\rm e}/m_{\rm e}c^2)^2$,
the characteristic time of \gr emission  decreases with energy
as $\propto E_\gamma^{-1/2}$. This may naturally explain less dramatic
variation of GeV \gr fluxes during the
keV/TeV  flares; the relatively low energy electrons
responsible for GeV radiation
(as well as for synchrotron optical/UV photons)
cannot promptly respond  to
changes of physical conditions in the jets. 
Very hard spectra of IC radiation below 100 GeV explain also the 
low flux of GeV \grs from Mkr~421, and their absence at all in the case of Mrk~501. 
%

%fig12ab (SAX)
\begin{wrapfigure}[26]{l}{7.5 cm}
\epsfxsize=7.5 cm
\epsffile[47 176 560 680]{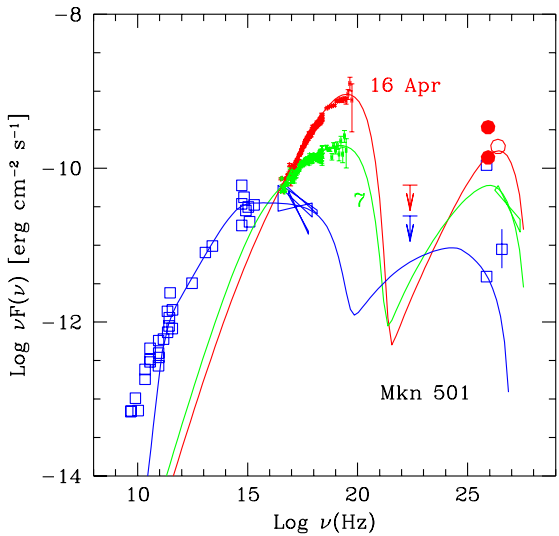}
\caption[]{Spectral energy distribution of Mrk~501 [94].  
BeppoSAX data from 7 and 16 April 1997
are indicated as labeled. Nearly simultaneous Whipple TeV data are indicated 
by filled circles [93], while the open circle 
(13 April) and TeV spectral fit (March 15-20)
are from the HEGRA [5]. Upper limits at 100~MeV are from [93] 
Non-simultaneous measurements collected 
from the literature are shown as open squares. The solid lines indicate fits with a 
one-zone, homogeneous synchrotron-self-Compton model.} 
\end{wrapfigure}

Although  there could be several external sources of seed photons
contributing to
the IC processes in AGN jets \cite{derm_schi,sikora},
the recent theoretical studies~\cite{maraschi,masti_kirk}
show that in Mrk~421 and Mrk~501 
the synchrotron photons  play a dominant role in production of
IC VHE $\gamma$-rays (the so-called  synchrotron-self-Compton (SSC) model). 
This reduces the number of model parameters,
and allows for conclusive predictions about the
broad-band spectral and temporal characteristics of radiation.
In practice, this  implies that the VHE radiation, being combined with
X-ray observations, is likely to be the most important window
in the electromagnetic spectrum  informing us
about nonthermal processes in  the BL Lac objects. 
This is demonstrated in Fig.~12 where the almost simultaneous spectral 
measurements of Mrk~501 flares in April 1997
by the  BeppoSAX satellite (over the range 1-200~keV) 
and by Whipple and HEGRA telescopes (at TeV energies) are presented together with 
the theoretical spectra calculated in the framework 
of homogeneous SSC model~\cite{mrk501_pian}.

VHE emission could be expected from many other
BL Lac objects. The sensitivity of future
IACT arrays $f_{\rm E} \sim 10^{-13} \, \rm erg/cm^2 s$ implies that all blazars 
emitting \grs at $E^{\prime} \geq 10 \, \rm GeV$ (in the frame of the jet) 
should be detected at energies $E \geq 100 \delta_{10} \, \rm GeV$ 
($\delta_{10}=\delta_{\rm j}/10$)
if the {\it intrinsic} \gr luminosity of the source exceeds 
$L^{\prime}=4 \pi d^2 f_{\rm E} /\delta_{\rm j}^4 
\simeq 2 \cdot 10^{38} \, (z/0.1)^2 \, \delta_{10}^{-4} \, \, \rm erg/s$ -- 
a rather small
amount of energy compared with the total energy budget 
(bolometric luminosity) of these most powerful
objects in the Universe !  

Note however  that, due to highly sporadic behavior of these objects,
the \gr emission may become visible only in short time episodes. Therefore
detection  of VHE outbursts from these objects with the small FoV 
Cherenkov telescopes require well justified search strategies. As an 
effective guide in such studies could be immediate information 
about the X-ray activity obtained on the daily basis
by the all-sky X-ray monitors like
ASM detector on RXTE satellite (R.Remillard, private communication).

\vspace{2mm}

\noindent
{\sc sources of highest energy cosmic rays.}~
There is little doubt
that the highest energy particles observed in CRs
up to $E \geq 10^{20} \, \rm eV$ are produced outside of the
Galaxy. Powerful nonthermal
objects like radiogalaxies, AGNs,
and clusters of galaxies~\cite{EHE_rev},
as well as more exotic objects,
e.g. enigmatic \gr burst sources \cite{EHE_GRB}, are all suggested
as possible sites of acceleration of these particles.
In many cases, these objects may contain sufficient target material
in the form of gas and photon fields to convert 
the energy of the parent particles into {\it detectable}
VHE radiation. A special interest may present the clusters of galaxies,
since the \grs produced at interactions of relativistic protons
with the ambient intracluster  gas carry important information about the
`cosmological' cosmic rays~\cite{vlk_clusters}. 
The efficiency of production
of $\pi^{0}$-decay \grs, due to the low density of the ambient gas in these objects 
$n \sim 10^{-5} - 10^{-4} \, \rm cm^{-3}$), 
is not high. However, the enormous energy in the relativistic protons,
accumulated over the life time of the Universe, $t=1/H_{0} \sim 10^{10} \, \rm yr$,
in the rich clusters of galaxies could be as high as $10^{62} - 10^{63} \, \rm erg$,
leading to fluxes which, in principle,  could be marginally detected
by 100 GeV threshold IACT arrays. 

More effective could be production of \grs due to interactions of 
protons with the ambient photon fields, provided
that the spectrum of accelerated particles extends beyond $10^{20} \, \rm eV$.  
Even if these particles do not spend
much time in their production region, but quickly leave the source,
they unavoidably collide with photons of 2.7~K MBR at relatively small
(on the cosmological scale) distances from the source (see Fig.~2).
The secondary electrons and photons -- the products of 
photo-meson reactions -- initiate pair cascades
in the 2.7~K MBR, leading
eventually to a flat standard ($dJ/dE \propto E^{-1.5}$)
\gr spectra  extending up to energy 100~TeV.
However, due to the further interaction with the extragalactic
infrared/optical photon fields, this
radiation arrives to the observer with a cutoff at much lower energies,
$E_{\rm cut}  \leq 10 \, \rm TeV$ (see Fig.~2).
Since the flux of \grs below $E_{\rm cut}$ is determined essentially
by the cascade processes, it is possible to give a model-independent
estimate of expected VHE fluxes:
$J(\geq 100 \, \rm GeV) \sim  10^{-12} (W_{\rm p}/10^{60} \, {\rm erg})
(d/1000 \, \rm Mpc)^{-2} \, \rm ph/cm^2 s$,
emitted within the
angle $\theta \sim l/d$ from an extended  region surrounding the
source of $E \geq 10^{20} \, \rm eV$ protons with total energy $W_{\rm p}$
The linear  size of the region $l$ emitting \grs
is limited by the
attenuation length of rectilinearly propagating
$\geq 10^{20} \, \rm eV$ protons
in the 2.7~K MBR, 
$l \sim \lambda_{p\gamma} \leq 100 (1+z)^{-3} \, \rm Mpc$.
Thus,  the angular size of the source does not significantly exceed
$1^{\circ}$, provided that the source is located at $d \geq 1000 \, \rm Mpc$.
In fact, the \gr production region could be
significantly smaller due to  non-rectilinear propagation of protons
in strong magnetic fields, which may exist in the several Mpc proximity
of extragalactic sources.
Since the nonthermal energy of many powerful extragalactic sources,
like radiogalaxy Cygnus A, could significantly exceed
$10^{60} \, \rm erg$, we may conclude that  100~GeV threshold
IACT arrays could be able to provide search for potential
accelerators of highest energy CRs up to distances $d \leq 1000 \, \rm Mpc$.
         
\vspace{2mm}

\noindent
{\sc gamma ray bursts.}~
The \gr bursts (GRBs) - accidental events of
impulsive radiation of soft (tens to several hundred keV) \grs
with duration from about $10^{-2} \, \rm s$ to $10^{3} \, \rm s$ -
represent an isotropically distributed source population of transient
astrophysical objects of unknown origin. 
One of the key aspects of this phenomenon -- the distance scale to the 
sources (``galactic or cosmological ?'') -- 
which since its discovery in 1973 was a topic of 
intense theoretical speculations,
finally seems to have been settled. 
After recent localization
by BeppoSAX the coordinates of 2 GRB events on Feb 28 and May 8 (1997) 
within a few arcminutes,
afterglows in X-ray, optical, and radio bands were discovered. The 
observed redshifted optical (absorption) lines with $z \geq 0.7$
is a strong evidence that the GRBs have cosmological origin, and
are located at the edge of the Universe (see e.g. [98] and references therein), 
which implies enormous energy, most probably in the nonthermal form, 
released on very short timescales.  
%
%fig13 (GRB)
\begin{wrapfigure}[22]{l}{6.8 cm}
\epsfxsize=6.8 cm
\epsffile[50 285 560 670]{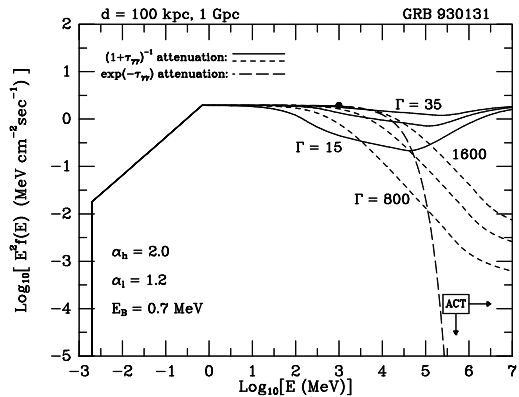}
\caption[]{The attenuation, internal to ther source, calculated for
of GRB 930131 detected al low energy \grs by BATSE at  
source distances typical of galactic halo 
($d=100 \, \rm kpc$, solid curves)
and cosmological origin ($d=1 \, \rm Gpc$, dashed curves), and different
bulk Lorentz factors $\Gamma$ indicated at the curves. 
The filled circle denotes the 
highest energy photon detected by EGRET. The current sensitivity 
of the Whipple telescope 
for observation of bursts is indicated by the ``ACT'' box''
[99].}
\end{wrapfigure}

This posses a challenge for explanation of the production and
escape of \grs from very compact regions, especially if 
one takes into account that
high energy $\gamma$-radiation above 100 MeV detected by EGRET from 
several GRBs, may be common for all bursts~\cite{baring_kp}. 
Actually, within the framework of the
most popular cosmological models of GRBs which assume that
the radiation is produced (independent of the nature of the energy source)
in the so-called relativistic fireballs~\cite{mesh_rees}, the \gr {\it production} spectrum 
could extend  beyond 100 GeV.
But this still does not imply VHE visibility of GRBs 
due to the internal $\gamma$-$\gamma$ absorption. The problem of  
self-opacueness  of the source 
can be effectively overcome~\cite{bar_hard} assuming 
bulk motion  with Lorentz factor $\Gamma \gg 1$, 
which actually is the  case of the relativistic fireballs.  
However, in cosmologically
distant GRBs even extremely large 
Lorentz factors, $\Gamma \geq 1000$, cannot prevent significant absorption of
VHE $\gamma$-rays (Fig.~13). Moreover, further strong attenuation
of VHE \grs takes place also in 
the intergalactic space~\cite{MHF}.
Whether or not the spectra of GRBs
continue to the VHE region is a question which remains one of the
interesting issues for ground-based \gr observations.
Irrespective of uncertainties of model parameters,
the fact of delayed high energy \gr emission, including
a 18~GeV photon detected by EGRET  after 1.5~h of
the famous February 17 event GRB940217 \cite{GRB_Fev17}
seems to be a warrant for a success provided that the
the energy threshold of detectors
would be reduced down to 20~GeV. Hopefully, the minimum 
detectable fluence of low-threshold Cherenkov detectors
estimated for a burst with duration $\Delta t$
as $S(\geq 20 \, {\rm GeV}) \sim 10^{-9}
(\Delta t/1 \, \rm s)^{1/2}  \rm erg/cm^2$
would allow detection of high energy tails of many GRBs.

\section{Observational Cosmology with VHE Gamma Rays}

The extension of the spectra of extragalactic sources to
beyond 100~GeV opens a new exciting aspect 
of ground-based gamma ray astronomy - {\it observational cosmology}.
The promise here is connected with {\it energy-dependent} 
interaction of VHE \grs with the diffuse extragalactic background
radiation (DEBRA). The absorption features in the spectra
of extragalactic sources~\cite{gould,steck1,primack}, as well as
secondary pair cascade radiation~\cite{prot_stan,halo,vhe_emis}
contain unique cosmological information about the 
intergalactic photon and magnetic  fields.

{\vspace{2mm}

\noindent
{\sc photon-photon absorption features}.~ 
An observer looking within a narrow cone centered on 
the source at a distance $d=c z/H_0$  will see an absorbed 
spectrum $J(E)=J_0(E) \exp (-\tau)$,
where $J_0(E)$ is the initial \gr spectrum, and $\tau(E,z)$ is the
optical depth to $\gamma$-$\gamma$ pair production 
on isotropically distributed photons of DEBRA. 
Except for the 2.7~K MBR our knowledge of density of the DEBRA 
today and its evolution in time, $u_\epsilon=\epsilon^2 n(\epsilon,z)$,
is rather  limited. This results in large uncertainties
in the optical depths (see Fig.~2 and Fig.~14). Therefore    
the study of the VHE \gr spectra from 
extragalactic objects  may provide important information about $u_\epsilon$ 
at different cosmological epochs.
Indeed, no deviation of the observed spectrum from the intrinsic 
(source) spectrum at energy $E$, e.g. by a factor of 
$\leq 2$ implies $\tau(E) \leq \rm ln2$. For a
low-redshift source ($z \ll 1$) this gives
(with accuracy $\leq 50 \%$ connected with  the uncertainty of the 
spectral shape of the DEBRA) 
an upper limit on the current density of DEBRA~\cite{GRO_Rev} at  
$\epsilon \approx 1 \,(E/ 1 \, \rm TeV)^{-1} \, \rm eV$: 
$u_{\epsilon} 
< 3.5 \cdot 10^{-3} \, (z/0.1)^{-1} \, (E/ 1 \, \rm TeV)^{-1} \,
(H_0/100 \, \rm km/s \, Mpc) \,  \rm eV/cm^3$.

Now, if one interprets the lack of an obvious   
cutoff in the measured $\gamma$-ray spectra of 
both Mrk~421  and Mrk~501 ($z \simeq 0.03$) up to $10 \, \rm TeV$ 
as an indication for a negligible absorption of $\gamma$-rays in the DEBRA,
and assuming  for the Hubble constant the maximum possible  value 
$H_0=100 \, \rm km/s \, Mpc$,
a conservative upper limit on $u_\epsilon$ could be 
derived from this simple, but model independent estimate:
$u_\epsilon \leq 1.1 \times 10^{-3} \, \rm eV/cm^3$. The recent 
detailed theoretical studies of the problem~\cite{stanev,stecker2},
based, in particular, on  realistic assumptions about 
the spectral shape of the DEBRA, give similar results. 
One of the interesting consequences of such a strong upper limit  
on the infrared  background could be an argument in favor of absence of 
strong evolution of galaxies  beyond $z=3$ (see Fig.~14). 

Formally, this upper limit can be extrapolated to shorter wavelengths
by assuming {\it a priori} power-law spectrum of DEBRA with differential
index $> 2$ (e.g. [111]), and thus to get more ``valuable'' (restrictive)
information about the DEBRA. But given the strong energy-dependence of the pair
production cross-section, this hardly can be justified. 
The model-independent constraint,
for example at 3 eV, could be derived  {\it only} from the lack of 
absorption feature at $E \sim 300 \, \rm GeV$. In the case of Mrk~421 and Mrk~501 
this
implies a rather high upper limit $u_\epsilon \leq 4 \times 10^{-2} \, \rm eV/cm^3$.
Deeper probe of the DEBRA at optical wavelengths is contingent only on the discovery
of distant ($z \geq 0.1$) VHE sources.

Obviously these interesting possibilities can be successfully utilized
only in the case of proper spectroscopic $\gamma$-ray measurements, as well as
good understanding of the intrinsic source spectra. The last condition seems to
be crucial, especially since the {\it cutoff} in a $\gamma$-ray spectrum 
does not yet unambiguously imply an {\it evidence for the intergalactic absorption},
and vice versa, the {\it lack of the cutoff}  
still does not automatically imply an {\it absence of intergalactic absorption}.
Interestingly, some  DEBRA models predict  a {\it modulation},  rather than cutoff ,
in the energy spectra of nearby AGNs ($d \sim 100 \, \rm Mpc$), at least
up to 10 TeV. Such modulation makes steeper the primary spectrum, but still
keeps it in the (quasi) power-law form (see Fig.~14).  
From this point of view the above stated upper limit
on the DEBRA based on the observed {\it power-law spectra} of Mrk~421
and Mrk~501  up to 10 TeV should be taken as rather circumstantial estimate.
Note that in the case of realization of the  
SSC models which assume a common population of parent electrons  producing
synchrotron X-rays and IC  $\gamma$-rays, the simultaneous keV and TeV 
observations of AGNs could provide an important, 
although not completely model-independent,
information about the intrinsic spectra of VHE $\gamma$-rays.

%
%fig14ab (DEBRA)
\begin{figure*}[t]
\vspace{6. cm}
\includegraphics{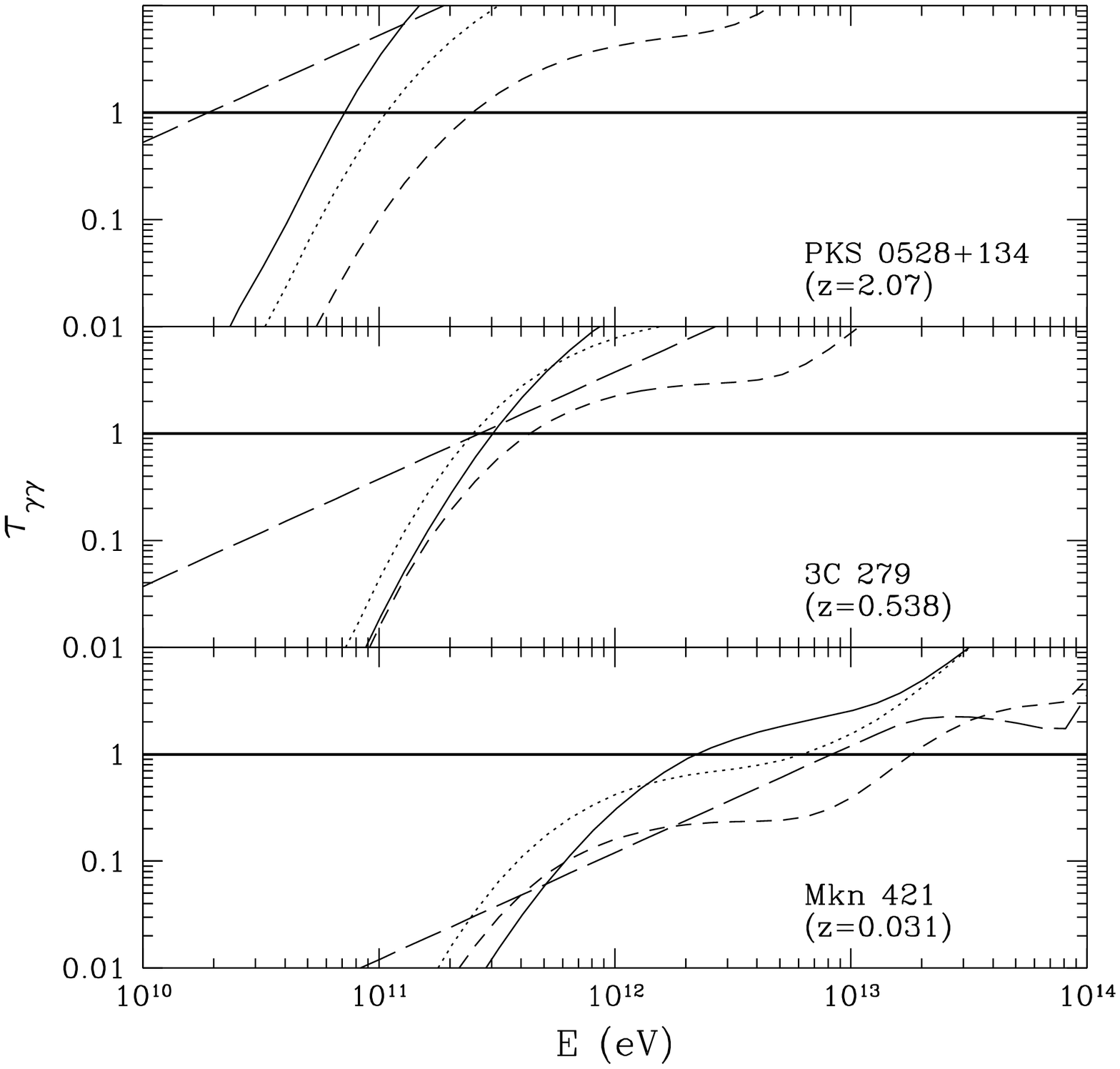}
\includegraphics{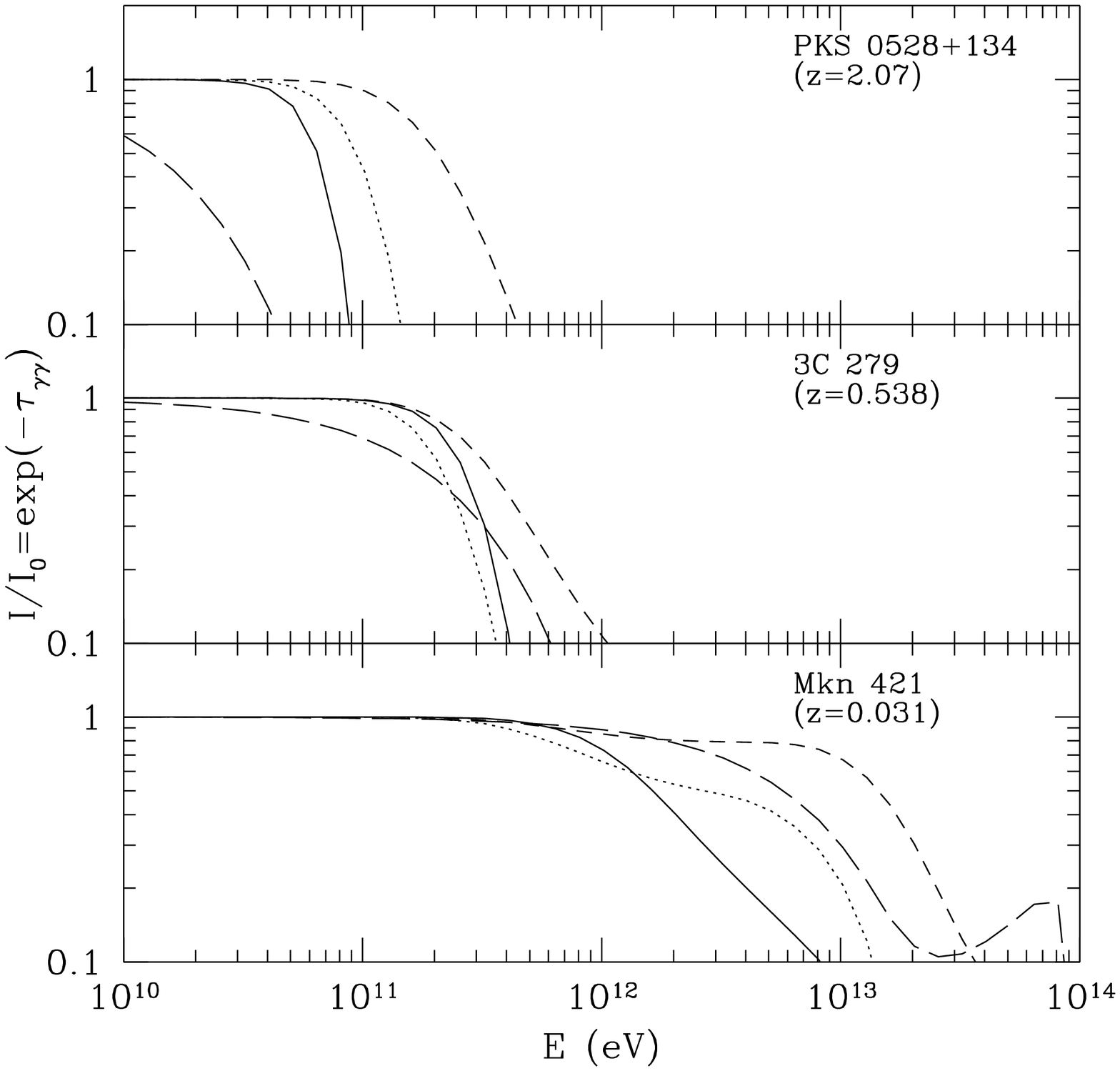}
\caption[]{Optical depth $\tau(E,z)$ {\bf (left)} 
and the spectrum attenuation factor $J(E)/J_0(E)=\exp(-\tau)$ {\bf (right)}
versus the energy for  $\gamma$-rays emitted
by close (Mrk~421), moderately distant (3C~273),
and very distant (PKS~0528+134) EGRET AGNs
for different cosmological scenarios of
formation and evolution of DEBRA  [37].
The solid line corresponds to the models of early ($z_{\rm form} \sim 5$)
formation of galaxies and stars in the Universe;
the short dashed line and dotted line
show the case of late ($z_{\rm form} \sim 1$) and
intermediate ($1 \leq z_{\rm form} \leq 3$)
stages of galaxy formation
which roughly correspond to so-called {\it Hot-Cold Dark Matter}
and {\it Cold Dark Matter} cosmological scenarios [106];
the long dashed curve
corresponds to the DEBRA density today, taken in the form
$n(\epsilon)=10^{-3} \epsilon^{-2} \, \rm ph/cm^3 eV$,
with a {\it naive} assumption about the
evolution of DEBRA back in redshifts as
$n(\epsilon) \propto (1+z)^3$.
}
\end{figure*}

\vspace{2mm}

\noindent
{\sc signatures of intergalactic pair cascades}.~~
The intergalactic absorption features, in the form of cutoffs or spectral
modulations, depend on the product $u_{\epsilon} \times H_0$, thus formally
they provide information about the Hubble constant as well \cite{Sal1}.
However, the measurements of the absorption features alone cannot decouple
$u_{\epsilon}$ and $H_0$. Separation of these two parameters requires more
{\it observables}. In principle, this can be done by studying
the spectral and  angular characteristics of VHE radiation from the presumable
pair halos \cite{halo} expected around the 
powerful extragalactic VHE sources
as a result of electromagnetic
cascades initiated by primary \grs in DEBRA. The radiation of the halo can be
recognized by its  distinct spectral and angular features which  only weakly
depend on the details of the central source. Therefore detection of a halo would give
us {\it two observables}, i.e. angular and spectral distributions of the radiation,
and thus would allow us to disentangle the two {\it variables}  $u_{\epsilon}$
and $H_0$. Remarkably, since the apparent angular size of the halo around a source
with known  redshift $z_0$ is determined by the level of DEBRA at the epoch
$z_0$, this allows an important study of the cosmological evolution of DEBRA
by observing pair halos from sources at different redshifts. Also, detections of both
the {\it primary} (point source-like and variable)  and the {\it halo}  (extended
and steady) components of VHE $\gamma$-ray emission from extragalactic sources allow
us to estimate the instantaneous and time-integrated VHE power of  sources,
and thus to measure the total nonthermal energy released by {\it individual}
extragalactic sources during their active phase (thought to be
$\sim 10^{6}-10^{8} \, \rm years$).

Obviously,  VHE radiation from pair halos  might be detectable around many more
objects than the ones presently active. The mapping and
spectroscopy of extended pair halos is a
difficult technical task, but it is feasible with the  new generation
100~GeV threshold IACT arrays, provided that
the total VHE luminosity of the central source at a distance $d$ exceeds 
$L_{\rm VHE}=10^{46} (d/1 \, \rm Gpc)^2 \, \rm erg/cm^2 s$ (see Fig.~15)
%

%fig15 (halo)
\begin{wrapfigure}[22]{l}{7.5 cm}
\epsfxsize=7.5 cm
\epsffile[90 250 490 620]{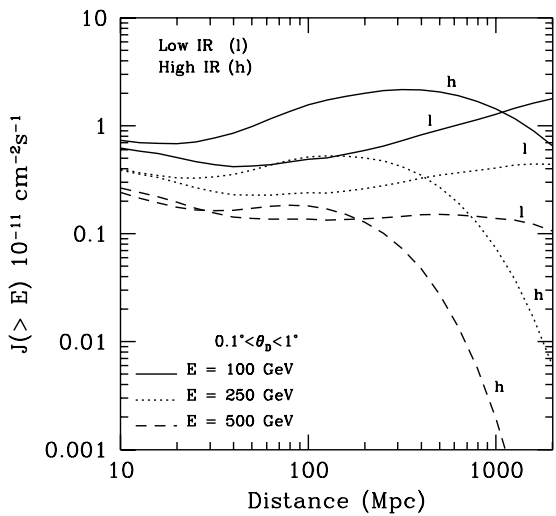}
\caption[]{Expected halo radiation fluxes 
integrated within $1^{\circ}$ above 100~GeV, 250~GeV, and 500 GeV
as a function of the source distance, calculated for two extreme (`Low' and `High')
levels of DEBRA. The assumed VHE luminosity of the central source above 10 TeV
is  $L_{\rm VHE}=10^{46} (d/1 \, \rm Gpc)^2 \, \rm erg/cm^2 s$.}
\end{wrapfigure}

The formation of isotropic pair halos around the extragalactic objects requires
intergalactic magnetic fields (IGMF) exceeding $10^{-12} \, \rm G$. At much
weaker IGMF, which formally cannot be excluded, instead of detection of
the extended  and persistent isotropic halos, one should expect cascade radiation
propagating  almost rectilinearly from the source to the observer. Even so,
the small deflections of the cascade electrons  by the extremely  weak magnetic
fields may induce noticeable time delays of the arriving high energy photons, as far
as cosmological distance scales are concerned. If such delays would  be possible to
distinguish  from the intrinsic time structure of a compact variable source,
this would a valuable  tool~\cite{Plaga} for a probe of primordial intergalactic
magnetic fields down to
$\leq 10^{-15} \, \rm eV$.

Irrespective to the structure and strength of IGMF, the ensemble of all VHE sources
in the Universe produce isotropic cascade $\gamma$-radiation. The spectrum of this
component  is rather insensitive to the spectrum of the primary VHE radiation.
Furthermore, at low energies, $E \leq 100 \, \rm GeV$, where the detection of
the isotropic component of $\gamma$-radiation are available by space-based 
$\gamma$-ray instruments,  the spectrum of the cascade radiation is only
weakly  sensitive also to the details of the spectrum and the 
flux of DEBRA. Therefore,
the  recent spectral measurements of the extragalactic $\gamma$-ray background by EGRET 
provide a robust constraint on the average VHE emissivity allowed in the Universe
in form of of any VHE phenomena 
(e.g. \gr emission of AGNs 
or decays of hypothetical  relic objects like topological 
defects~{\cite{vhe_emis,TD_prdt,TD_sigl}) 
comparing the calculated cascade  \gr background flux accumulated over the Hubble time with
the measured diffuse flux at GeV energies~\cite{vhe_emis}: 
$L_{\rm VHE} \leq (1-3) \times 10^{50} \, \rm erg/s$.  
In fact this 
upper limit could be interpreted also as a {\it measure}
of VHE luminosity of the Universe as a whole, if one assumes~\cite{Pohl} 
that the hard EGRET  spectrum  is explained by the cascade component, 
initiated in the intergalactic medium  by VHE \grs from AGNs,  
rather than by direct  contribution of  $\gamma$-ray
emitting AGNs to the extragalactic diffuse background.

\vspace{2mm}

In conclusion, the recent results of 
ground-based gamma ray astronomy as well as exciting  
theoretical/phenomenological predictions show that 
the Universe likely to be well  populated by
a variety of TeV \gr emitters. The further study of 
the sky in VHE \grs by forthcoming powerful
instruments promises a new path 
towards the understanding of the nonthermal 
high energy phenomena in the Universe.

\section*{Acknowledgments}

The invitation of the organizers of the XVIII International Symposium
on ``Lepton-Photon Interactions'' is greatly appreciated. I am grateful to
many my colleagues from the `TeV Gamma Ray Astronomy'
community for providing materials during the preparation of this paper. 
Special thanks should go to my collaborators A.M. Atoyan, P.S. Coppi, 
A.K. Konopelko and H.J. V\"olk for many aspects of the work
summarized here.


\section*{References}
\begin{thebibliography}{99}

{\small

%%%%%%%%%%%%%%%%%%%%%%%%%%%%%%%%%%%%%%%%%%%%%%%%%%%%%%%%%%%%%%%%%%%%%%%%%% SEC.1 %%%%%%%%%%%%
%1
\bibitem{egret}C.E. Fichtel {\it et al,} {\it Astrophys.J. Suppl.} 
{\bf 90}, 917 (1994). 
%2
\bibitem{glast}E.D. Bloom, {\it Space Sci. Rev.} {\bf 75}, 109 (1996).
%3
\bibitem{dingus}B.L. Dingus,
{\it Proc. 25th ICRC} (Durban) {\bf 5}, 69 (1997). 
%4
\bibitem{whipple_crab97}G. Mohanty {\it et al,} {\it Astropart. Physics},
submitted (1997).
%5
\bibitem{hegra_mkn501_AA}F.A. Aharonian {\it et al,} 
{\it Astron. Astrophys.}
{\bf 327}, L5 (1997).
%6
\bibitem{cang_crab}T. Tanimori {\it et al,} {\it Astrophys.J.},
in press (1997). 
%7
\bibitem{casa_crab97}A. Borione {\it et al,} {\it Astrophys.J.} 
{\bf 481}, 313 (1997). 
%8
\bibitem{atah_crab/mnras}A.M. Atoyan and 
F.A. Aharonian, {\it Mon. Not. Royal  
Astron. Soc.} {\bf 278}, 525 (1996).
%9
\bibitem{integral}P. von Balmoos,  
{\it Space Sci. Rev.} {\bf 75}, 83 (1996).
%%%%%%%%%%%%%%%%%%%%%%%%%%%%%%%%%%%%%%%%%%%%%%%%%%%%%%%%%%%%%%%%%%%%%%%%%%% SEC.2 %%%%%%%%%
%10
\bibitem{cronin_annrev93}J.W. Cronin, K.G. Gibbs and  T.C. Weekes,
{\it Annual Rev. Nucl. Part. Sci.} {\bf 43}, 883 (1993). 
%11
\bibitem{whipple_crab89}T.C. Weekes {\it et al,} {\it Astrophys.J.} 
{\bf 342}, 379 (1989). 
%12 
\bibitem{Caw/Weekes_96}M.F. Cawley and  T.C. Weekes,
{\it Experimental Astronomy} {\bf 6}, 7 (1996).
%13
\bibitem{AhAk_annrev97}F.A. Aharonian and C.W. Akerlof, 
{\it Annual Rev. Nucl. Part. Sci.} {\bf 47}, 273 (1997).
%14
\bibitem{Weekes_Turver}T.C. Weekes and  K.E. Turver, in {\it Proc. 12th
ESLAB Symposium} (Frascati), 279 (1977).
%15
\bibitem{Stepanian}A.A.Stepanian {\it et al,} {\it Izv. Krym. Astrophyz. Obs.}
{\bf 66}, 234 (1983).
%16
\bibitem{Hillas_LaJolla}A.M. Hillas, {\it Proc 19th ICRC} 
(La Jolla) {\bf 3}, 445 (1985). 
%17
\bibitem{GRO_Rev}T.C. Weekes, F.A.Aharonian, D.J.Fegan and T.Kifune, in 
{\it Proc. 4th Compton Symposium}, (eds. C.D.  Dermer, M.S. Strickman, 
and J.D. Kurfess), AIP Conf. Proc. 410,  361 (1997). 
%18
\bibitem{cang_vela}T. Yashikoshi  {\it et al,} {\it Astrophys.J.}, 
submitted (1997).
%19 
\bibitem{cang_sn1006}T. Tanimori {\it et al,} {\it Astrophys.J.}, 
submitted (1997).
%20
\bibitem{whipple_1es}M. Catanese {\it et al,} {\it Proc. 25th ICRC} 
(Durban) {\bf 3}, 277 (1997).
%21
\bibitem{Weekes_Pysrep88}T.C. Weekes, {\it Physics Reports} 
{\bf 160}, 1 (1988)
%22
\bibitem{Fegan_ad90}D.J. Fegan, {\it Proc. 25th ICRC} (Adelaide) 
{\bf 11}, 23 (1990).
%23
\bibitem{potch_AEaq}P.J. Meintjes {\it et al,} {\it Astrophys.J.}
{\bf 434}, 292 (1994).
%24
\bibitem{durham_AEaq}P.M. Chadwick {\it et al,} {\it Astropart. Physics}
{\bf 4}, 99 (1995).
%25
\bibitem{Perpig_hegra}F.A. Aharonian and G.Heizelmann, 
{\em Nucl. Phys.} B, in press (1997).
%26
\bibitem{potch_VelaX-1}B.C. Raubenheimer {\it et al,} {\it Astrophys.J.}
{\bf 428}, 777 (1994).
%27 
\bibitem{durham_cenx3}P.M. Chadwick {\it et al,} in preparation (1997).
%28
\bibitem{egret_cenx3}T.W. Vestrand and P. Sreekumar, in 
{\it Proc. 4th Compton Symposium}, (eds. C.D.  Dermer, M.S. Strickman,
and J.D. Kurfess), AIP Conf. Proc.10 (1997).
%29
\bibitem{buckley_SNR}J.H. Buckley {\it et al,}
{\it Astron. Astrophys.}, in press (1997).
%30
\bibitem{SNRrev_GRO4}O.C. de Jager and M.G.Baring, in 
{\it Proc. 4th Compton Symposium}, (eds. C.D.  Dermer, M.S. Strickman, 
and J.D. Kurfess), AIP Conf. Proc.10 (1997). 
%31
\bibitem{vlk_SNR}H.J. V\"olk, in preparation (1997)
%32
\bibitem{Cronin_Rome}J.W. Cronin, {\it IL Nuovo Cimento} {\bf 19C},   
847 (1996).
%33
\bibitem{casa_GALdiffuse}A. Borione {\it et al,} 
{\it Astrophys.J}, {\bf 493}, in press (1998).
%34
\bibitem{casa_EXGdiffuse}M.C. Chantell {\it et al,} 
{\it Phys. Rev. Lett} {\bf 79}, 1805 (1997). 
%35
\bibitem{Ces_Paris92}C.J. Cesarsky, {\it Nucl. Phys.} B {\bf 28B}, 51 (1992).
%36
\bibitem{Hillas_annrev84}A.M. Hillas, {\it Ann. Rev. Astron.Astrophys.} 
{\bf 22}, 425 (1984).
%37
\bibitem{vhe_emis}P.S. Coppi and F.A. Aharonian,  {\it Astrophys.J. Lett.}
{\bf 487}, L9 (1997).
%38
\bibitem{milagro}G.B. Yodh, {\it Space Sci. Rev.} {\bf 75}, 199 (1996).
%39
\bibitem{large_zenith}P. Somers and J.W. Elbert, 
{\it J. Phys. G: Nucl.Phys.} {\bf 13}, 553 (1987).
%%%%%%%%%%%%%%%%%%%%%%%%%%%%%%%%%%%%%%%%%%%%%%%%%%%%%%%%%%%%%%%%%%%%%%%%%%%% Sec.3 %%%%%%%%%%%
%40
\bibitem{CAT}P. Goret, 
{\it Proc. 25th ICRC} (Durban) {\bf 3}, 173 (1997).  
%41
\bibitem{Hillas_HD}A.M. Hillas,  {\it Space Sci. Rev.} {\bf 75}, 17 (1996).
%42
\bibitem{Fegan_97}D.J. Fegan, {\it J.Phys.. G: Nucl. Part. Phys.} {\bf 23},
1013 (1997).
%43
\bibitem{Granite_Lamb}R.C. Lamb  {\it et al,} in {\it Towards a Major
Atmospheric Cherenkov Detector-IV} (Padova), ed. M.Cresti, 386 (1995).
%44
\bibitem{W_Mkn421flare}J.A.Gaidos {\it et al,} {\it Nature} 
{\bf 383} 319 (1996).
%45
\bibitem{Daum_Crab}A. Daum {\it et al,} {\it Astropart. Physics},
in press (1997).
%46
\bibitem{hegra_perf}F.A. Aharonian, in
{\it Proc. 4th Compton Symposium}, (eds. C.D.  Dermer, M.S. Strickman,
and J.D. Kurfess), AIP Conf. Proc. 410, I. 1397; II. 1641 (1997).
%47
\bibitem{TA_mkn501}M. Teshima {\it et al,} in preparation (1997). 
%48
\bibitem{Whipple_dish}M.F. Cawley {\it et al,} 
{\it Experimental Astronomy} {\bf 1}, 173 (1990).
%49
\bibitem{CAT_camera}M. Punch {\it et al,} in {\it Towards a Major
Atmospheric Cherenkov Detector-IV} (Padova), ed. M. Cresti, 356 (1995).
%50
\bibitem{HEGRA_camera}G. Hermann {\it et al,} in {\it Towards a Major
Atmospheric Cherenkov Detector-IV} (Padova), ed. M. Cresti, 396 (1995).
%51
\bibitem{Array}F.A. Aharonian {\it et al,} {\it Astropart. Physics}
I. {\bf 6}, 343; II. {\bf 6}, 369 (1997).
%52
\bibitem{TArray_1992}M. Teshima {\it et al,} in {\it Towards a Major
Atmospheric Cherenkov Detector-I} (Palaiseau), eds. P.Fleury and 
G. Vacanti, 255 (1992).
%53
\bibitem{veritas}T.C. Weekes {\it et al,} Proposal submitted
to the Smithsonian Astrophys. Obs. (1996).
%54
\bibitem{hess}F.A. Aharonian {\it et al,} Letter of Intent,
MPI f\"ur Kerphysik, Heidelberg (1997).
%55
\bibitem{Lamb_Snowmass}R.C. Lamb {\it et al,} in {\it Particle and
Nuclear Astrophysics and Cosmology in the Next Millennium},
ed. E.W. Kolb and R.D. Peccei (World Scientific, Singapore), 295 (1995).
%56
\bibitem{Magic}S.M. Bradbury   {\it et al,} in {\it Towards a Major
Atmospheric Cherenkov Detector-IV} (Padova), ed. M. Cresti, 277 (1995).
%57
\bibitem{STACEE}R.A. Ong {\it et al.} 
{\it Astroparticle Physics} {\bf 5}, 353 (1996).
%58
\bibitem{CELESTE}J. Quebert {\it et al,} 
{\it Proc. 25th ICRC} (Durban) {\bf 5}, 145 (1997).
%%%%%%%%%%%%%%%%%%%%%%%%%%%%%%%%%%%%%%%%%%%%%%%%%%%%%%%%%%%%%%%%%%%%%%% SEC.4 %%%%%
%59
\bibitem{BBDGP}V.S. Berezinsky {\it et al,} 
{\it Astrophysics of Cosmic Rays} (North-Holand, 1990).
%60
\bibitem{Ramana}P.V. Ramana Murthy and A.W. Wolfendale,
{\it Gamma Ray Astronomy} (Cambridge, 1993).
%61
\bibitem{DAV}L.O'C. Drury, F.A. Aharonian F A and 
H.J. V\"olk,  {\it Astron.Astrophys.} {\bf 287}, 959 (1994).
%62
\bibitem{Naito}T. Naito and F. Takahara, 
{\it J.Phys.. G: Nucl. Part. Phys.} {\bf 20}, 477 (1994).
%63
\bibitem{Berezhko}E.G. Berezhko and H.J. V\"olk,
{\it Astroparticle Physics} {\bf 7}, 183 (1997).
%64
\bibitem{ASCA_1006}F. Koyama {\it et al,} {\it Nature} {\bf 378}, 255 (1995)
%65
\bibitem{masti_1006}A. Mastichiadis and O.C. de Jager,
{\it Astron. Astrophys.} {\bf 311}, L5 (1996).
%66
\bibitem{Ber_Ptus}V.S.Berezinskii and V.S.Ptuskin, 
{\it Sov. Astron. Lett.} {\bf 13}, 304 (1988).
%67
\bibitem{KDB_youngsnr}J.G. Kirk, P. Duffy and  L. Ball,
{\it Astron. Astrophys.} {\bf 293}, L37 (1995).
%68
\bibitem{ahat_clouds}A.A. Aharonian and  A.M. Atoyan,
{\it Astron. Astrophys.} {\bf 309}, 917 (1996).
%69
\bibitem{egret_galdif}S.D. Hunter {\it et al,}
{\it Astrophys.J.} {\bf 481}, 205 (1980).
%70
\bibitem{electrons_nish}J.Nishimura {\it et al,}
{\it Astrophys.J.} {\bf 238}, 394 (1980).
%71
\bibitem{EGRT_puls}P.L. Nolan {\it et al,}
{\it Astron. Astrophys. (Suppl.)} {\bf 120}, 61 (1996).
%72
\bibitem{Harding_puls}A.K. Harding, {\it Astr. Soc. Pacific Conf. Ser.},
on press (1997).
%73
\bibitem{Romani}R. Romani,
{\it Astrophys.J.} {\bf 470}, 469 (1996).
%74
\bibitem{Arons}J. Arons,  {\it Space Sci. Rev.} {\bf 75}, 235 (1996).
%75
\bibitem{Harding_DH}A.K. Harding, {\it Space Sci. Rev.} {\bf 75}, 257 (1996).
%76
\bibitem{ahatkif_mnras}F.A. Aharonian, A.M. Atoyan and T. Kifune,
{\it Mon. Not. Royal  Astron. Soc.} {\bf 291}, 162 (1997).
%77
\bibitem{Xnebula_asca}K. Tamura {\it ae al,} {\it Publ. Astron. Soc. Japan}
{\bf 48}, L33 (1996).
%78
\bibitem{cang_1706}T. Kifune {\it ae al,} 
{\it Astrophys.J.} {\bf 438}, L91 (1996).
%79
\bibitem{von_Mont}C. von Montigny {\it et al,} 
{\it Astrophys. J.} {\bf 440}, 525 (1995).
%80
\bibitem{urry} C.M. Urry and  P. Padovani, {\it PASP} {\bf 107}, 803 (1995).
%81
\bibitem{mannheim}K. Mannheim, {\it Astron. Astrophys.} {\bf 269}, 67 (1993).
%82
\bibitem{dar_agn}A. Dar and  A. Laor, {\it Astrophys.J.} {\bf 478}, L5 (1997).
%83
\bibitem{derm_schi} C.D. Dermer and R. Schlickeiser, 
{\it Astrophys. J.} {\bf 415}, 458 (1993).
%84
\bibitem{sikora}M. Sikora, M.C. Begelman, and M.J. Rees,
{\it Astrophys. J.} {\bf 421}, 153 (1994).
%85
\bibitem{marscher}S.D. Bloom and A.P. Marscher, 
{\it Astrophys. J.} {\bf 461}, 657 (1996).
%86
\bibitem{maraschi}G. Ghisellini, L. Maraschi and L. Dondi,
{\it Astron. Astrophys. (Suppl.)} {\bf 120}, 503 (1996).
%87
\bibitem{takahara}S. Inoue and F. Takahara, {\it Astrophys. J.} 
{\bf 463}, 555 (1996).
%88
\bibitem{masti_kirk}A. Mastichiadis and J.G. Kirk,
{\it Astron. Astrophys.} {\bf 320}, 19 (1997).
%89
\bibitem{bed_prot}W. Bednarek and  R.J. Protheroe,
{\it Mon. Not. Royal  Astron. Soc.} {\bf 287}, L9 (1997).
%90
\bibitem{ahat_beam/target}F.A. Aharonian and A.M. Atoyan,
{\it Space Sci. Rev.} {\bf 75}, 357 (1996).
%91
\bibitem{mkr421_asca}T. Takahashi T {\it et al,}
{\it Astrophys. J.} {\bf 470}, L89 (1996).
%92
\bibitem{mkr421_whipple}J.H. Buckley {\it et al,} 
{\it Astrophys. J.} {\bf 472}, L9 (1996). 
%93
\bibitem{mkr501_catanese}M. Catanese {\it et al,}
{\it Astrophys. J.} {\bf 487}, L143 (1997).
%94
\bibitem{mrk501_pian}E. Pian {\it et al,}
{\it Astrophys. J.}, in press (1997).
%95
\bibitem{EHE_rev}P. Sokolsky, P. Sommers and B.R. Dawson,
{\it Phys. Rep.} {\bf 217}, 225 (1992).
%96 
\bibitem{EHE_GRB}E. Waxman, {\it Phys. Rev. Lett.} {\bf 75}, 386 (1995).
%97
\bibitem{vlk_clusters}H.J. V\"olk, F.A. Aharonian and Breitschwerdt,
{\it Space Sci. Rev.} {\bf 75}, 279 (1996).
%98
\bibitem{GRB_news}J.S. Bloom and M. Ruderman, {\it Nature} {\bf 387}, 859 (1997).
%99 
\bibitem{baring_kp}M.G. Baring, in preparation (1997).
%100
\bibitem{mesh_rees}P. Meszaros and  M.J. Rees, 
{\it Astrophys. J.} {\bf 476}, 232 (1997).
%101
\bibitem{bar_hard}M.G. Baring and A.K. Harding,
{\it Astrophys. J.} {\bf 481}, L85 (1997).
%102
\bibitem{MHF}K. Mannheim, D. Hartmann and B. Funk,
{\it Astrophys. J.} {\bf 467}, 532 (1996).
%103
\bibitem{GRB_Fev17}K. Hurley, {\it Nature} {\bf 372}, 652 (1994).
%%%%%%%%%%%%%%%%%%%%%%%%%%%%%%%%%%%%%%%%%%%%%%%%%%%%%%%%%%%%%%%%%%%%%% SEC. 5 %%%%%
%104
\bibitem{gould}R.J. Gould and G. P. E. Schreder, 
{\it Phys. Rev. Lett.} {\bf 16}, 252 (1966). 
%105
\bibitem{steck1}F.W. Stecker, O.C. de Jager and M.H. Salamon,
{\it Astrophys. J.}, {\bf 390}, L49 (1992).
%106
\bibitem{primack}D. Macminn and J.R. Primack, 
{\it Space Sci. Rev.} {\bf 75}, 413 (1996).
%107
\bibitem{prot_stan}R.J. Protheroe and T. Stanev,
{\it Mon. Not. Royal  Astron. Soc.} {\bf 264}, 191 (1993).
%108
\bibitem{halo}F.A.  Aharonian, P.S. Coppi and H.J. V\"olk, 
{\it Astrophys. J.}, {\bf 423}, L5 (1994).
%109
\bibitem{stanev}T. Stanev and A. Franceschini,
{\it Astrophys. J.}, submitted (1997).
%110
\bibitem{stecker2}F.W. Stecker and O.C. de Jager,
{\it Astrophys. J.}, submitted (1997).
%111
\bibitem{DeJag}O.C. Jager, M.H. Salamon and 
F.W. Stecker, {\it Nature} {\bf 369}, 294 (1994).
%112
\bibitem{Sal1} M.H. Salamon, F.W. Stecker and O.C. de Jager O. C., 
{\it Astrophys. J.} {\bf 423}, L1 (1994). 
%113
\bibitem{Plaga}R. Plaga, {\it Nature} {\bf 374}, 430 (1995).
%114
\bibitem{TD_prdt}R. Protheroe and T.Stanev, {\it Phys. Rev. Lett.} 
{\bf 77}, 3708 (1996). 
%115
\bibitem{TD_sigl}G. Sigl, S. Lee, D.N. Schramm  
and  P.S. Coppi, {\it Phys. Lett.} B {\bf 392}, 129 (1997). 
%116
\bibitem{Pohl}M. Pohl {\it et al,}
{\it Astron. Astrophys.} {\bf 326}, 51 (1997). 
}

\end{thebibliography}
\end{document}